\RequirePackage{fix-cm}
\documentclass[epj,nopacs]{svjour}   
\smartqed  
\RequirePackage{graphicx}
\RequirePackage{mathptmx}      
\RequirePackage{flushend}
\RequirePackage[numbers,sort&compress]{natbib}
\RequirePackage[colorlinks,citecolor=blue,urlcolor=blue,linkcolor=blue]{hyperref}
\usepackage{revsymb}

\usepackage{multirow}             
\usepackage{tabularx}             
\usepackage{booktabs}             
\usepackage{amsmath}              
\usepackage{bm}

\newcommand{\AuAu}    {Au\,+\,Au collisions at 1.23\agev}   
\newcommand{\agev}    {\mbox{$A$~GeV}}               
\newcommand{\gevc}    {\mbox{GeV$/c$}}

\newcommand{\mev}     {\mbox{MeV}}
\newcommand{\mevc}    {\mbox{MeV$/c$}}

\newcommand{\rb}[1]   {\mbox{\textrm{\scriptsize #1}}}

\newcommand{\pimin}   {\ensuremath{\pi^{-}}}
\newcommand{\piplus}  {\ensuremath{\pi^{+}}}

\newcommand{\linv}    {\ensuremath{\lambda_{\rb{inv}}}}
\newcommand{\losl}    {\ensuremath{\lambda_{\rb{osl}}}}
\newcommand{\rinv}    {\ensuremath{R_{\rb{inv}}}}

\newcommand{\qinv}    {\ensuremath{q_{\rb{inv}}}}
\newcommand{\qout}    {\ensuremath{q_{\rb{out}}}}
\newcommand{\qside}   {\ensuremath{q_{\rb{side}}}}
\newcommand{\qlong}   {\ensuremath{q_{\rb{long}}}}

\newcommand{\rout}    {\ensuremath{R_{\rb{out}}}}
\newcommand{\rside}   {\ensuremath{R_{\rb{side}}}}
\newcommand{\rlong}   {\ensuremath{R_{\rb{long}}}}
\newcommand{\routlong}   {\ensuremath{R_{\rb{out\,long}}}}
\begin{document}
\title{Identical pion intensity interferometry at $\boldmath{\sqrt{s_{\mathrm{NN}}}=}$2.4~GeV}
%
\titlerunning{Identical pion intensity interferometry at $\sqrt{s_{\mathrm{NN}}}=$2.4~GeV} 
\author{HADES collaboration \\[5bp]
J.~Adamczewski-Musch$^{4}$, O.~Arnold$^{10,9}$, C.~Behnke$^{8}$, A.~Belounnas$^{16}$,
A.~Belyaev$^{7}$, J.C.~Berger-Chen$^{10,9}$, J.~Biernat$^{3}$, A.~Blanco$^{2}$, C.~~Blume$^{8}$,
M.~B\"{o}hmer$^{10}$
% ...
}
\institute{
\mbox{ } \\[-8bp]
\mbox{$^{1}$Institute of Physics, Slovak Academy of Sciences, 84228~Bratislava, Slovakia}\\
\mbox{$^{2}$LIP-Laborat\'{o}rio de Instrumenta\c{c}\~{a}o e F\'{\i}sica Experimental de Part\'{\i}culas , 3004-516~Coimbra, Portugal}\\
\mbox{$^{3}$Smoluchowski Institute of Physics, Jagiellonian University of Cracow, 30-059~Krak\'{o}w, Poland}\\
\mbox{$^{4}$GSI Helmholtzzentrum f\"{u}r Schwerionenforschung GmbH, 64291~Darmstadt, Germany}\\
\mbox{$^{5}$Technische Universit\"{a}t Darmstadt, 64289~Darmstadt, Germany}\\
\mbox{$^{6}$Institut f\"{u}r Strahlenphysik, Helmholtz-Zentrum Dresden-Rossendorf, 01314~Dresden, Germany}\\
\mbox{$^{7}$Joint Institute of Nuclear Research, 141980~Dubna, Russia}\\
\mbox{$^{8}$Institut f\"{u}r Kernphysik, Goethe-Universit\"{a}t, 60438 ~Frankfurt, Germany}\\
\mbox{$^{9}$Excellence Cluster 'Origin and Structure of the Universe' , 85748~Garching, Germany}\\
\mbox{$^{10}$Physik Department E62, Technische Universit\"{a}t M\"{u}nchen, 85748~Garching, Germany}\\
\mbox{$^{11}$II.Physikalisches Institut, Justus Liebig Universit\"{a}t Giessen, 35392~Giessen, Germany}\\
\mbox{$^{12}$Institute for Nuclear Research, Russian Academy of Science, 117312~Moscow, Russia}\\
\mbox{$^{13}$Institute of Theoretical and Experimental Physics, 117218~Moscow, Russia}\\
\mbox{$^{14}$National Research Nuclear University MEPhI (Moscow Engineering Physics Institute), 115409~Moscow, Russia}\\
\mbox{$^{15}$Department of Physics, University of Cyprus, 1678~Nicosia, Cyprus}\\
\mbox{$^{16}$Institut de Physique Nucl\'{e}aire, CNRS-IN2P3, Univ. Paris-Sud, Universit\'{e} Paris-Saclay, F-91406~Orsay Cedex, France}\\
\mbox{$^{17}$Nuclear Physics Institute, The Czech Academy of Sciences, 25068~Rez, Czech Republic}\\
\mbox{$^{18}$LabCAF. F. F\'{\i}sica, Univ. de Santiago de Compostela, 15706~Santiago de Compostela, Spain}\\
\mbox{$^{19}$Uniwersytet Warszawski - Instytut Fizyki Do\'{s}wiadczalnej, 02-093 Warszawa, Poland}\\
\\[-5bp]
\mbox{$^{a}$ also at Coimbra Polytechnic - ISEC, ~Coimbra, Portugal}\\
\mbox{$^{b}$ also at ExtreMe Matter Institute EMMI, 64291~Darmstadt, Germany}\\
\mbox{$^{c}$ also at Technische Universit\"{a}t Dresden, 01062~Dresden, Germany}\\
\mbox{$^{d}$ also at Frederick University, 1036~Nicosia, Cyprus}\\
\mbox{$^{e}$ also at Dipartimento di Fisica and INFN, Universit\`{a} di Torino, 10125~Torino, Italy}
\\ $^{\dag}$ Deceased.
}
\authorrunning{J.~Adamczewski-Musch et al.} 
\date{Received: \today}
%
\abstract{
High-statistics $\pi^-\pi^-$ and $\pi^+\pi^+$ femtoscopy data are presented for 
Au\,+\,Au collisions at $\sqrt{s_\mathrm{NN}} = 2.4$~GeV, 
measured with HADES at SIS18/GSI. The experimental correlation functions allow the 
determination of the space-time extent of the corresponding emission sources 
via a comparison to models. 
The emission source, parametrized as three-dimensional Gaussian distribution, 
is studied in dependence on pair transverse momentum, azimuthal 
emission angle with respect to the reaction plane, collision centrality and beam energy. 
For all centralities and transverse momenta, a geometrical distribution of ellipsoidal shape 
is found in the plane perpendicular to the beam direction with the larger extension 
perpendicular to the reaction plane. For large transverse momenta, the corresponding   
eccentricity approaches the initial eccentricity. The eccentricity is smallest 
for most central collisions, where the shape is almost circular. The magnitude of the 
tilt angle of the emission ellipsoid in the reaction plane decreases with 
increasing centrality and increasing transverse momentum. 
All source radii increase with centrality, largely exhibiting a 
linear rise with the cube root of the number of participants. 
A substantial charge-sign difference of the source radii is found,  
appearing most pronounced at low transverse momentum. The extracted 
source parameters are consistent with the extrapolation of their energy dependence down 
from higher energies.
%
\PACS{{}25.75.-q, 25.75.Gz}
%
} 
\maketitle
%
%
\section{Introduction}
\label{sect:intro}
Two-particle intensity interferometry of hadrons is widely used to study
the spatio-temporal size, shape and evolution of their source in
heavy-ion collisions or other reactions involving hadrons 
(for a review see ref.\,\cite{Lisa05}). 
The technique, pioneered by Hanbury Brown and Twiss \cite{HBT56} to measure angular 
radii of stars, later on named HBT interferometry, is based on the quantum-statistical 
interference of identical particles. 
Goldhaber et al. \cite{GGLP60} first applied intensity interferometry to hadrons.   
In heavy-ion collisions,  
the intensity interferometry does not allow to measure directly the reaction volume,
as the emission zone, changing shape and size in the course of the collision, 
is affected by dynamically generated
space-momentum correlations ({\it e.g.} radial expansion after the compression phase 
or resonance decays).
Thus, intensity interferometry generally does not yield the proper source size, but rather
an effective ``length of homogeneity'' \cite{Lisa05}. It measures source regions in which 
particle pairs are close in momentum, so that they are correlated as a consequence of 
their quantum statistics or due to their two-body interaction. At energies in the GeV region, 
the measured particles can originate from many different processes. 
Therefore, the intensity interferometry may 
provide additional information to the understanding of reaction mechanisms which finally 
determine the particle emission sources. 

In general, the sign and strength of the correlation is affected by (i) the
strong interaction, (ii) the Coulomb interaction if charged particles
are involved, and (iii) the quantum statistics in the case of identical
particles (Fermi-Dirac suppression for fermions, Bose-Einstein enhancement for bosons). 
In the case of $\pi\pi$ correlations, the mutual strong interaction appears to be 
negligible \cite{Bowler88} compared to the effects (ii) and (iii). 

It is worth emphasizing that, before the measurement presented here, 
only preliminary data \cite{hbt_fopi_1995} of  
identical-pion HBT data exist for a large symmetric collision system (like Au\,+\,Au or 
Pb\,+\,Pb) at a beam kinetic energy of about $1\agev$ (fixed target, 
$\sqrt{s_\mathrm{NN}} = 2.3$~GeV)\footnote{Throughout this publication 
$\agev$ refers to the mean kinetic beam energy.}. 
For the somewhat smaller system La\,+\,La, studied at 
$1.2A$~GeV with the HISS spectrometer at the Lawrence Berkeley Laboratory (LBL) Bevalac, 
pion correlation data were reported by Christie et al. \cite{Christie93,Christie92}. 
An oblate shape of the pion source and a correlation of the source size with the system 
size were found. Also, pion intensity interferometry for small systems (Ar\,+\,KCl, 
Ne\,+\,NaF) was studied at 1.8$A$~GeV at the LBL Bevalac using the Janus spectrometer by 
Zajc et al. \cite{Zajc84}. Both groups made first attempts to correct the influence of 
the pion-nuclear Coulomb interaction on the pion momenta. The effect on the source radii, 
however, was found negligible for their experiments. 

In this article we report on the investigation of $\pi^-\pi^-$ and $\pi^+\pi^+$  
correlations at low relative momenta in \AuAu~ 
(fixed target, $p_\mathrm{proj}/A=1.96$~\gevc, $\sqrt{s_\mathrm{NN}} = 2.41$~GeV), 
continuing our previous femtoscopic studies of smaller collisions systems 
\cite{hades_pLambda_ArKCl,hades_hbt_ArKCl,hades_pLambda_pNb}. In a recent letter  
\cite{hades_HBT_letter} presenting only the results of the azimuthally-integrated HBT analysis 
for central collision, we reported a substantial charge-sign 
difference of the source radii, particularly pronounced at low transverse momenta, 
and a smooth extrapolation of the $\sqrt{s_\mathrm{NN}}$ dependence of the source parameters 
towards low energies. Here, we present the complete HBT analysis, including  
azimuthal-angle and centrality dependences.   
In Sect.\,\ref{sect:experiment} we shortly describe the
experiment. In Sect.\,\ref{sect:CF} we define the correlation function and discuss possible  
distortions to it. In Sect.\,\ref{sect:results}
we present the three-dimensional pion emission source resulting from the 
correlation analysis, its dependences on collisions centrality and kinematic quantities 
and compare our observations to the findings of other experiments. Finally, 
we summarize our results in Sect.\,\ref{sect:summary} and give an outlook.


\section{The experiment}
\label{sect:experiment}
The measurement was performed 
with the \textbf{H}igh \textbf{A}cceptance \textbf{D}i-\textbf{E}lectron \textbf{S}pectrometer 
(HADES) at the Schwerionensynchrotron SIS18 at GSI, Darmstadt. HADES, although
primarily optimized to measure di-electrons \cite{HADES-PRL07}, offers also excellent
hadron identification capabilities
\cite{hades_kpm_phi_arkcl,hades_xi_arkcl,hades_K0_ArKCl,hades_Lambda_ArKCl,hades_xi_pNb,hades_kpm_phi_auau}.
The setup of the HADES experiment is described in detail in
\cite{Agakishiev:2009am}.  HADES is a charged particle detector
consisting of a six-coil toroidal magnet centered around the beam axis
and six identical detection sections located between the coils and
covering polar angles between $18^{\circ}$ and $85^{\circ}$.  Each
sector is equipped with a Ring-Imaging Cherenkov (RICH) detector
followed by four layers of Mini-Drift Chambers (MDCs), two in
front of and two behind the magnetic field, as well as a scintillator
Time-Of-Flight detector (TOF) ($44^{\circ}$~--~$85^{\circ}$) and
Resistive Plate Chambers (RPC) ($18^{\circ}$~--~$45^{\circ}$).  
TOF, RPC, and Pre-Shower detectors (behind RPC, for e$^\pm$ identification) 
were combined into a Multiplicity and Electron Trigger Array (META). 
Charged hadron identification is based on the time-of-flight measured with TOF and
RPC, and on the energy-loss information from TOF as well as from the
MDC tracking chambers.  Electron candidates are in addition selected
via their signals in the RICH detector. Combining this information
with the momentum, $p$, as determined from the deflection of the 
tracks in the magnetic field, allows to identify charged particles
(e.g. pions, kaons or protons) with a high significance. Particles are assumed to  
be identified as pions if their velocity, $\beta$, is found within a $3\sigma$ window 
around the theoretical expectation, $\beta_\pi=p/\sqrt{p^2+m_\pi^2}$. The  
corresponding cut parameters $\sigma$ are derived from Gaussian fits to the velocity 
distribution in slices of momentum. (We use units with $\hbar=c^2=1$.)  

Several triggers are implemented.  The
minimum bias trigger is defined by a signal in a diamond START detector
in front of the 15-fold segmented gold target.  
In addition, online Physics Triggers (PT) are used, which
are based on hardware thresholds on the TOF signals, proportional to
the event multiplicity, corresponding to at least 5 (PT2) or 20 (PT3)
hits in the TOF.  Events are selected offline by requiring that their
global event vertex is inside the target region, i.e. within an interval of 
65~mm along the beam axis. About 2.1 billion 
Au\,+\,Au collisions corresponding to the 43\,\% most central events 
are taken into account for the present correlation analysis. 

The centrality determination is based on the summed number of hits 
detected by the TOF and the RPC detectors. The measured events are 
divided in centrality classes corresponding to 
intervals of the integrated cross section. 
Note that, as a result of the analogue hardware threshold of the online 
trigger, the most peripheral class can be extended to 45\,\% (or even further), 
i.e. beyond the above given effective value.

The respective average impact parameters and mean numbers of 
nucleons participating in the formation of the nuclear fireball, 
$\langle A_\mathrm{part} \rangle$, as deduced from extensive Monte-Carlo (MC) 
calculations \cite{hades_centrality:2018am,PhobosGlauber2015} 
with the Glauber model \cite{Glauber:1955qq} are 
summarized in table\,\ref{tab:rp_resolution}.  

The determination of the reaction plane angle, $\phi_{\mathrm{RP}}$, is based 
on the measurement of the charged projectile spectator fragments (mostly $Z=1,\,2$) 
by their position, flight time and energy deposit. 
They are detected by a three square meter scintillator hodoscope  
7\,m downstream the target, consisting of 288 scintillator cells. 
It covers polar angles from 0.3 to 7.2 degrees. Due to the dispersion of the 
event plane a resolution correction needs to be applied (cf. Sect.\,\ref{sect:ana_azi_sens}). 

Finally, we want to note that, 
as result of both the neutron excess of the collision system and the different 
detector acceptances of negatively and positively charged particles, 
about five to six times more $\pi^-\pi^-$ pairs than $\pi^+\pi^+$ pairs are measured.   
\begin{table}[ht!]
\centering
\caption{
The mean number of participants (2nd column) and the average impact parameter (3rd column) 
calculated \cite{hades_centrality:2018am} 
with Glauber MC simulations \cite{PhobosGlauber2015} 
for the centrality classes listed in the 1st column. 
The 4th and 5th columns give the corresponding 1st and 2nd order reaction plane 
resolutions (cf.\,Sect.\,\ref{sect:ana_azi_sens}), 
respectively, calculated according to ref.\,\cite{Ollitrault97}.
  } 
\label{tab:rp_resolution}
 \begin{tabular}{c | c c c c} 
 \toprule 
 Centrality  & $\langle A_\mathrm{part} \rangle$ & $\langle b \rangle$ & $F_1$ & $F_2$ \\ 
      ($\%$) &                                   &      (fm)           &       &       \\ 
 \midrule 
  \phantom{0}$ 0\,-\,10$ & $303$ & $3.1$ & 0.648 & 0.298 \\
             $10\,-\,20$ & $213$ & $5.7$ & 0.847 & 0.572 \\
             $20\,-\,30$ & $150$ & $7.4$ & 0.887 & 0.653 \\
             $25\,-\,35$ & $125$ & $8.1$ & 0.886 & 0.651 \\
             $30\,-\,40$ & $103$ & $8.7$ & 0.876 & 0.629 \\
             $30\,-\,45$ & $~93$ & $9.0$ & 0.871 & 0.620 \\
 \bottomrule 
 \end{tabular}
\end{table}
\section{The correlation function }
\label{sect:CF}
Generally, the two-particle correlation function is defined as the ratio of the 
probability $P_2({\bm p}_1,{\bm p}_2)$ to measure simultaneously two particles with 
momenta ${\bm p}_1$ and ${\bm p}_2$ and the product of the corresponding single-particle 
probabilities $P_1({\bm p}_1)$ and $P_1({\bm p}_2)$ \cite{Lisa05}, 
\begin{equation}
C({\bm p}_1, {\bm p}_2) = \frac{P_2({\bm p}_1,{\bm p}_2)}{P_1({\bm p}_1)  P_1({\bm p}_2)}.
\label{def_theo_corr_fct}
\end{equation}
Experimentally this correlation is formed as a function of the momentum difference between the 
two particles of a given pair and quantified by taking the ratio of the yields of 'true' pairs 
($Y_\mathrm{true}$) and uncorrelated pairs ($Y_\mathrm{mix}$). $Y_\mathrm{true}$ is constructed  
from all particle pairs in the selected phase space interval from the same event.
$Y_\mathrm{mix}$ is generated by event mixing, where 
particle 1 and particle 2 are taken from different events. Care was taken to mix particles from 
similar event classes in terms of multiplicity, vertex position and reaction plane angle.
The events are allowed to differ by not more than 10 in the number of the RPC\,+\,TOF hit 
multiplicity (i.e. corresponding to the typical multiplicity uncertainty as deduced from 
simulations for fixed impact parameter \cite{hades_centrality:2018am}), 
1.2\,mm in the $z$-vertex coordinate (amounting to less than one third of the spacing between 
target segments), and 10 degrees in azimuthal angle relative to the reaction plane 
(significantly below the event plane 
resolution, cf. table\,\ref{tab:rp_resolution}), respectively. 
The momentum difference is decomposed into three orthogonal components as suggested by 
Podgoretsky~\cite{Podgoretsky83}, Pratt~\cite{Pratt86} and Bertsch~\cite{Bertsch89}.  
The three-dimensional correlation functions are projections of 
Eq.\,(\ref{def_theo_corr_fct}) into the (out, side, long)-coordinate system, 
where `out' means along the pair transverse momentum, 
${\bm p}_\mathrm{t,\,12} = {\bm p}_\mathrm{t,\,1}+ {\bm p}_\mathrm{t,\,2}$, `long' is 
parallel to the beam direction $z$, and `side' is oriented perpendicular to the other two 
directions. The particles forming a pair are boosted into the longitudinally comoving  
system, where the $z$-components of the momenta cancel each other, $p_{z_{1}}+p_{z_{2}}=0$. 
This system choice allows for an adequate comparison with correlation data taken at very 
different, usually much higher, collision energies, where the distribution of the rapidity, 
$y=\tanh^{-1}{(\beta_\mathrm{z})}$, of produced particles is found to be not as narrow as in 
the present case but largely elongated. (Here, $\beta_\mathrm{z}=p_\mathrm{z}/E$, 
$E=\sqrt{p^2+m_0^2}$ and $m_0$ are the longitudinal velocity, 
the total energy and the rest mass of the particle, respectively. )
Hence, the experimental correlation function is given by  
\begin{equation}
C(\qout,\qside,\qlong) = {\cal N} \,\frac{Y_{\mathrm{true}}(\qout,\qside,\qlong)}{Y_{\mathrm{mix}}(\qout,\qside,\qlong)},  
\label{def_exp_corr_fct_3dim}
\end{equation}
where $q_i=(p_\mathrm{1,\,i}-p_\mathrm{2,\,i})/2$ ($i$\,=\,'out',\,'side',\,'long') 
are the relative momentum 
components, and ${\cal N}$ is a normalization factor which is fixed by the requirement 
$C \rightarrow 1$ at large relative momenta, where the
correlation function is expected to flatten out at unity. The statistical errors of 
$C$ are dominated by those of the true yield, 
since the mixed yield is generated with much higher statistics. 
Analogously to Eq.\,(\ref{def_exp_corr_fct_3dim}), the experimental one-dimensional 
correlation function, 
\begin{equation}
C(\qinv) = {\cal N}' \,\frac{Y_{\mathrm{true}}(\qinv)}{Y_{\mathrm{mix}}(\qinv)},   
\label{def_exp_corr_fct_1dim}
\end{equation}
is generated by projecting Eq.\,(\ref{def_theo_corr_fct}) 
onto the Lorentz-invariant relative momentum, 
\begin{equation}\label{q_inv} 
\qinv =\frac{1}{2}\sqrt{({\bm p}_1 - {\bm p}_2)^2-(E_1-E_2)^2},  
\end{equation}
where $E_i=\sqrt{p^2_i+m^2_i}$ and $m_i$ ($i=1,\,2$) are the total energies 
and the rest masses of the particles forming the pair, respectively. 
Note that for two particles of equal mass, $\qinv$ is identical in magnitude 
to the single particle momenta in the rest frame of the pair. 
\subsection{Treatment of close-track effects}
\label{sect:eff_corr}
Two different methods to correct for the possible bias due to close-track effects introduced 
by the limitations of the HADES detector + track finding procedures are investigated and found 
to agree within statistical fluctuations.   

Method A, called double-ratio method, is based on UrQMD \cite{UrQMD} transport 
model simulations of \AuAu~fully transported 
through the HADES implementation HGeant of the 'Detector Description and 
Simulation Tool' GEANT3.21 \cite{GEANT}. 
Since UrQMD does not incorporate quantum-statistical or Coulomb effects, any identical-pion 
correlation function generated from the corresponding 
HGeant output according to Eq.\,(\ref{def_exp_corr_fct_1dim}) is expected to be flat at unity. 
Deviations from that value are considered to result from close-track effects. 
Hence, dividing the experimental and the simulated correlation functions should 
allow for a reasonable correction of the former one. However, this method is statistically 
limited because of the present availability to us of only 100 million UrQMD events, 
i.e. 20 times less than the experimental data sample. Thus, the statistical errors of the 
corrected correlation function would be dominated by those of the simulation, a 
drawback becoming particularly important when investigating the correlation 
function in a multi-dimensional parameter space. 
Also, it could not be fully guaranteed that HGeant is realistic enough in reproducing 
the behaviour of nearby tracks.  These simulations also showed that there are no significant  
long-range correlations, usually attributed either to energy-momentum conservation in 
correlation analyses of small systems or to minijet-like phenomena at high energies. 
Therefore, this method serves for cross checks but is not 
involved in the results presented in Sect. \ref{sect:results}.   

Method B implements appropriate selection conditions on the META-hit and MDC-layer level, 
i.e. by discarding pairs which hit the same META cell, and by excluding for particle\,2 
three successive wires symmetrically around the MDC wire fired by particle\,1. 
This method was tested with simulations based on UrQMD\,+\,HGeant and a detailed 
description of the detector response, to firmly exclude any close-track effect. 
Also broader exclusion windows have been investigated, but no significant 
improvement was found.  Though there is a certain amount 
of pairs with small relative momenta getting lost due to this condition 
(about $50\,\%$ for $\qinv<40~\mevc$), its superior statistical significance still  
clearly favors this method over the double-ratio method.    
Consequently, Method B, applied to both the true and the mixed-event yields, 
is used throughout the analysis \cite{RG} presented in this article. 
\subsection{Parameterization of one-dimensional correlation functions}
\label{sect:CF_qinv}
The fits to the one-dimensional correlation function are performed with the function    
\begin{equation}
C(\qinv) = N \left[1-\linv + \linv K_\mathrm{C}(\qinv,\rinv) C_\mathrm{qs}(\qinv)\right], 
\label{pipi_fit_fct}
\end{equation} 
using a Gaussian function for the quantum-statistical (Bose-Einstein) part, 
\begin{equation}
C_\mathrm{qs}(\qinv)=1+\exp{(-(2\qinv\rinv)^2)}.     
\label{pipi_fit_be}
\end{equation}
The influence of the mutual Coulomb interaction in Eq.\,(\ref{pipi_fit_fct}) is separated  
from the Bose-Einstein part by including in the fits the commonly used Coulomb correction 
by Sinyukov et al.\,\cite{Sinyukov98}.
The Coulomb factor $K_\mathrm{C}$ results from the integration of the 
two-pion Coulomb wave function squared over a spherical Gaussian source of fixed radius.   
This radius is iteratively approximated by the result of the corresponding fit to the 
correlation function. The parameters $N$ and $\lambda$ in Eq.\,(\ref{pipi_fit_fct}) 
represent a normalization constant and the fraction of correlated pairs, respectively. 

All fits performed to one-dimensional and three-dimensional (cf. Sect.\,\ref{sect:CF_qosl}) 
correlation functions involve a log-likelihood minimization \cite{Ahle02}. No 
significant differences are observed when using a $\chi^2$ minimization. 
\subsection{Parameterization of three-dimensional correlation functions}
\label{sect:CF_qosl}
The three-dimensional experimental correlation function is fitted with the function
\begin{align}
& C(\qout,\,\qside,\,\qlong) = \nonumber \\ 
& N \big[1-\losl  + \losl K_\mathrm{C}(\hat q,\rinv) C_\mathrm{qs}(\qout,\,\qside,\,\qlong)\big], 
\label{pipi_fit_fct_3dim}
\end{align} 
where, for azimuthally-integrated analyses (Sect.\,\ref{sect:ana_azi_int}) 
at midrapidity ($y_\mathrm{cm}=0.74$),   
\begin{align}
& C_\mathrm{qs}(\qout,\,\qside,\,\qlong)=  \nonumber \\ 
& 1+\exp{(-(2\qout\rout)^2-(2\qside\rside)^2-(2\qlong\rlong)^2)} \hfill
\label{pipi_fit_be_3dim}
\end{align} 
represents the quantum-statistical part of the correlation function and 
\begin{align}
\hat q= \qinv(\qout,\,\qside,\,\qlong,\bar k_\mathrm{t}) 
\label{qhat}
\end{align}
is the average value of the invariant momentum difference 
for given intervals of the relative momentum components and $k_\mathrm{t}$. 
Actually, Eq.\,(\ref{pipi_fit_be_3dim}) can be written in a more general way,  
\begin{align}
C_\mathrm{qs} = 1+\exp{(-4 \sum_{i,\,j} q_i R^2_{ij} q_j)},   
\label{pipi_fit_be_3dim_general}
\end{align} 
as used for the azimuthally-sensitive HBT analyses presented in Sect.\,\ref{sect:ana_azi_sens}.
For symmetry reasons \cite{UHeinz02} 
the non-diagonal ($i \ne j$) elements of $R_{ij}$, comprising 
the combinations 'out'-'side' and 'side'-'long', vanish when azimuthally and rapidity integrated  
analyses are performed \cite{e895_2000,HBT_in_UrQMD}, 
as is done in Sect.\,\ref{sect:ana_azi_int}. The 'out'-'long' component, however, 
should have a finite value depending on the degree of symmetry 
of the detector-accepted rapidity distribution w.r.t. midrapidity. 
We studied this effect by including in Eq.\,(\ref{pipi_fit_be_3dim}) 
an additional term $-2 \qout (2 \routlong)^2 \qlong$, where the prefactor accounts for both 
non-diagonal terms, 'out-long' and 'long-out'. 
We found only marginal differences in the fits which resulted, for all centrality and 
transverse-momentum classes, in rather small values of $R^2_\mathrm{out\,long}< 1$\,fm$^2$, 
which in many cases are consistent with zero within uncertainties.  
This finding is not surprising, since the rapidity 
distribution of charged pions accepted by HADES is almost centered at midrapidity and rather 
narrow, i.e. extending just over about $\pm 0.7$ rapidity units. 

Finally, for all results presented in Sect.\,\ref{sect:results}, we restricted 
the pair rapidity to an interval $\vert y - y_\mathrm{cm} \vert<0.35$, 
within which $dN/dy$ does not vary by more than 10\,\%. 
\subsection{Momentum resolution correction}
\label{sect:mom_res_corr}
The effect of the finite momentum resolution of the HADES tracking system 
\cite{Agakishiev:2009am} is studied 
with dedicated simulations. For that purpose, pairs of identical pions (resulting from the 
two-body decay of a fictive mother particle with its mass chosen to produce relative momenta of interest, e.g. $\qinv=10,\,20,\,30,\,...~\mevc$) are 
simulated with the event generator Pluto~\cite{Pluto} and subsequently tracked through 
HGeant, the latter one modelling the HADES detector with its granularity and 
momentum resolution. The relative-momentum distribution of the pion pairs delivered by the 
simulation is fitted with a Gaussian function. The resulting widths $\sigma_q(q_\mathrm{i})$ 
(i='inv',\,'out',\,'side',\,'long') are folded into the fit 
functions (Eqs.\,(\ref{pipi_fit_fct}) and (\ref{pipi_fit_fct_3dim})) to account for the slight 
resolution-induced decrease of both $R$ and $\lambda$.  A typical resolution amounts to  
$\sigma_q(\qinv=20~\mevc) \simeq 2~\mevc$ corresponding to a radius shift of 
$\Delta R/R \simeq +2$\,\% after the correction.
\subsection{Systematic error estimate}
\label{sect:syst_err}
The main contribution to the systematic uncertainties of the results presented 
in the subsequent section is due to the fluctuation of the fit results when varying the 
fit range over the respective relative-momentum quantity. 
{The standard fit range was chosen to an interval extending from 4 to $80~\mevc$. 
We varied the lower limit to $8~\mevc$ and the upper one from 60 to $100~\mevc$.
Typical radius changes of $0.1 - 0.3$\,fm are observed.

The contribution of the close-track effects discussed in Sect.\,\ref{sect:eff_corr} 
was estimated by varying the size of the wire-exclusion window (3 vs. 5 wires) in Method B. 
The resulting radius uncertainties did not exceed 0.2\,fm; 
typically they were smaller than 0.1\,fm. 

The influence of possible impurities entering the charged pion samples was tested with 
stronger cuts on the quality parameters of particle identification 
(cf. Sect.\,\ref{sect:experiment}), i.e. by a $\pm30~\mev$ mass window around the most 
probable pion mass. About two third of the pairs survived this cut. 
No systematic differences w.r.t. to the full data sample are found within the statistical errors.  

The effect of varying $R_{\mathrm{inv}}$ in the Coulomb correction in 
Eq.\,(\ref{pipi_fit_fct_3dim}) results in systematic uncertainties of $\sim 0.01$\, fm; 
the effect of the finite size of the averaging intervals 
in Eq.\,(\ref{qhat}) yields systematic uncertainties of even 
smaller size. 

The uncertainty of the momentum resolution correction described in Sect.\,\ref{sect:mom_res_corr} 
appears to be an order of magnitude smaller than the absolute source radius shift, 
i.e. typical values of $0.01 - 0.03$\,fm are considered. 

Another systematic uncertainty was estimated from studies of the forward-backward symmetry 
of the fit results w.r.t. midrapidity. Selecting the rapidity windows  
$-0.35<y-y_\mathrm{cm}<0$ and $0<y-y_\mathrm{cm}<0.35$  
and taking similar (unbiased by the detector acceptance) transverse momentum intervals, typical  
systematic variations of the fit radii of $0.03 - 0.1$\,(0.2)\,fm for 
$R_{\mathrm{inv}}$, $R_{\mathrm{side}}$, $R_{\mathrm{long}}$ ($R_{\mathrm{out}}$) are observed. 

For the fit of the azimuthally integrated three-dimensional correlation function, 
the slight differences of the results when switching on/off the 
'out'-'long' component in the fit function (cf. Sect.\,\ref{sect:CF_qosl}) 
are taken as further systematic uncertainty. Typical values of these differences 
are $0.03 - 0.15$\,fm.
 
As an additional cross check, the stability of the results w.r.t. a reversed setting 
(for about 10\,\% of the beam time) of the magnetic field has been investigated. 
Within the 
larger statistical errors, the results for $\pi^-\pi^-$ ($\pi^+\pi^+$) in reversed field are 
found identical to the $\pi^-\pi^-$ ($\pi^+\pi^+$) results in default field,  
although oppositely charged pions are subjected to different overall detector acceptances. 

Finally, all systematic error contributions are added quadratically. 
%
%
\section{Results}
\label{sect:results}
\subsection{Experimental correlation functions}
\label{sect:corr_fct_pimpim}
The data are divided into centrality classes (cf. Sect.\,\ref{sect:experiment}) 
and into classes of pair transverse momentum, $p_\mathrm{t,\,12}$, from which the 
pair transverse mass, $m_\mathrm{t}=\sqrt{k_\mathrm{t}^2+m_{\pi}^2}$ with 
$k_t=p_\mathrm{t,\,12}/2$, is derived. 
As width of the underlying $p_\mathrm{t,\,12}$ bins 100\,$\mevc$ was selected. 
In case of azimuthally-sensitive HBT analyses (Sect.\,\ref{sect:ana_azi_sens}), 
the bin size of the pair angle w.r.t. the reaction plane, $\Phi=\phi_{12}-\phi_\mathrm{RP}$,    
is chosen to be $\pi/4$. 

A representative one-dimensional $\pimin\pimin$ correlation function fitted with 
Eq.\,(\ref{pipi_fit_fct}) is shown in Fig.\,\ref{fig:Cinv_centr0to10_pt100to200}.
Note that we usually exclude from the fits 
pairs with $\qinv < 6~\mevc$, since at very small relative momenta 
slight remnants of close-track effects could not completely be 
avoided when applying the correction Method B described in Sect.\,\ref{sect:eff_corr}. 
This exclusion from the fit also renders a potential correction of 
the center of gravity of the first $\qinv$ bin unnecessary, within which the Coulomb 
factor $K_\mathrm{C}(\qinv)$ exhibits a rather steep increase. (However, already 
within the 2nd bin function $K_\mathrm{C}$ is fairly smooth. Moreover, since both the 
true and mixed yields naturally show different slopes at low $\qinv$, a unique positioning 
of the corresponding bin centers required by Eq.\,(\ref{def_exp_corr_fct_1dim}) 
would be quite sophisticated.) 
For the quality of the three-dimensional fits we refer to \cite{hades_HBT_letter}.

Two-dimensional projections of the Coulomb-corrected three-dimensional correlation function 
(Eq.\,(\ref{pipi_fit_fct_3dim})) are presented in 
Fig.\,\ref{fig:Cosl_centr10to30_pt100to400_phi0} for $\vert \Phi \vert<\pi/8$. 
The low-momentum enhancement due to the quantum-statistical (Bose-Einstein) effect 
is visible in all three directions. Comparing these in-plane 
correlations with the corresponding 
out-of-plane ($\vert\Phi-\pi/2\vert<\pi/8$) correlations displayed in  
Fig.\,\ref{fig:Cosl_centr10to30_pt100to400_phi90}, 
a modification of the $q$ range of the Bose-Einstein part of the correlation becomes visible. 
(Due to the permutability of particles 1 and 2, one of the $q$ projections 
can be restricted to positive values.) 
Detailed results of the azimuthally-dependent correlation functions are presented in 
Sect.\,\ref{sect:ana_azi_sens}. 

\begin{figure}[h]
\begin{center}
\includegraphics[width=0.9\linewidth]{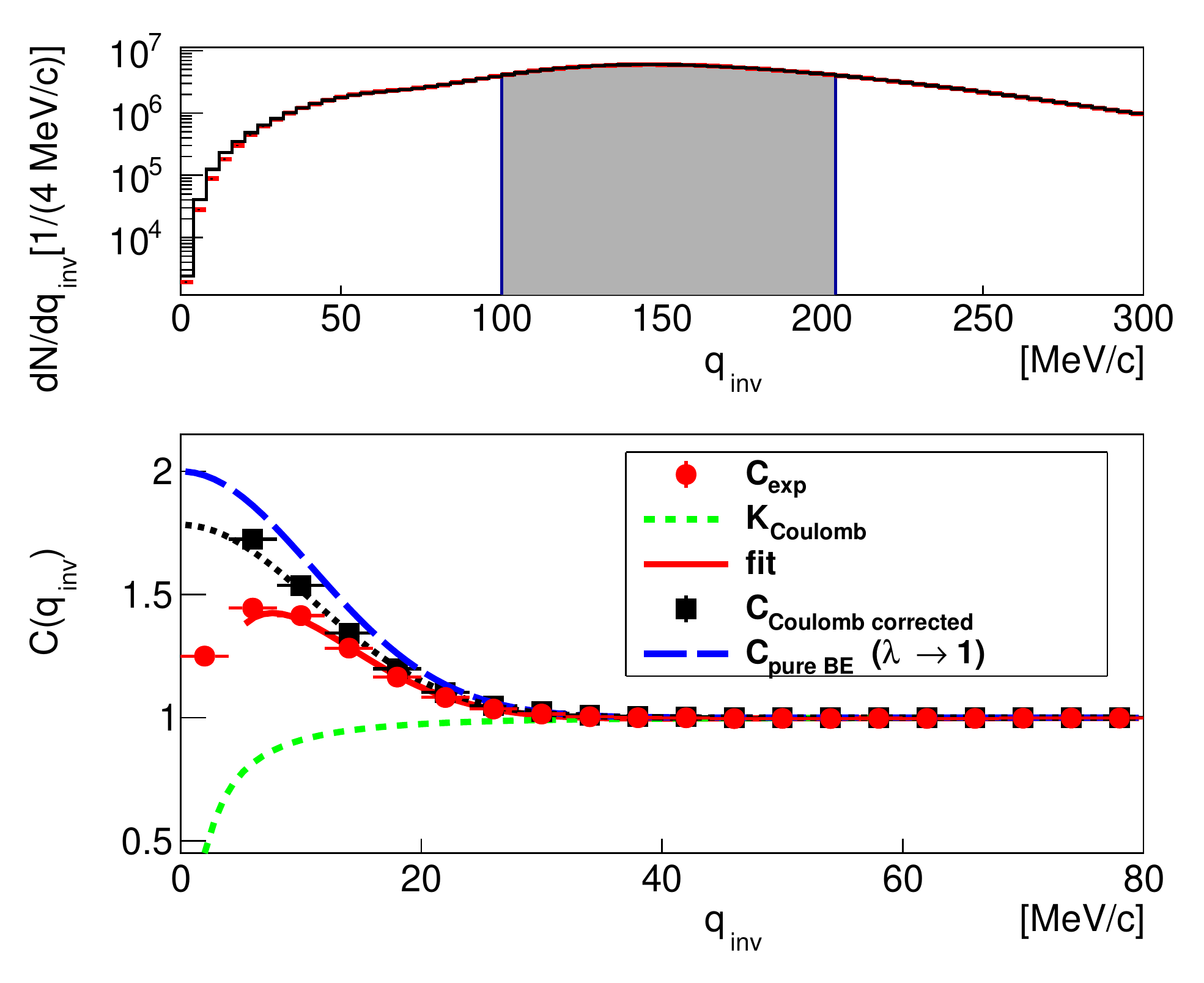}
\end{center}
\caption{Upper panel: The distribution of the invariant relative momentum $\qinv$ for 
$\pi^-\pi^-$ pairs with transverse momentum of $p_\mathrm{t,\,12}=100-200~\mevc$ 
for central ($0-10\,\%$) \AuAu. 
The black (red) histogram displays the true (mixed) yield. The grey-shaded area represents the 
yield used for normalization. 
Lower panel: The one-dimensional $\pi^-\pi^-$ correlation function as function of $\qinv$. 
Red circles display the ratio of the true and mixed distributions, respectively.  The green 
dashed curve represents the Coulomb correction function $K_C$ as described in 
Sect.\,\ref{sect:CF_qinv}. The black squares correspond to 
the Coulomb-corrected correlation function. 
The red full (black dotted) curve shows the fit function (Eq.\,(\ref{pipi_fit_fct})) before 
(after) the Coulomb correction. The blue long-dashed curve gives the pure Bose-Einstein part 
(Eq.\,(\ref{pipi_fit_be})) of the correlation function. }
\label{fig:Cinv_centr0to10_pt100to200}
\end{figure}
\begin{figure}[h]
\begin{center}
\includegraphics[width=0.9\linewidth]{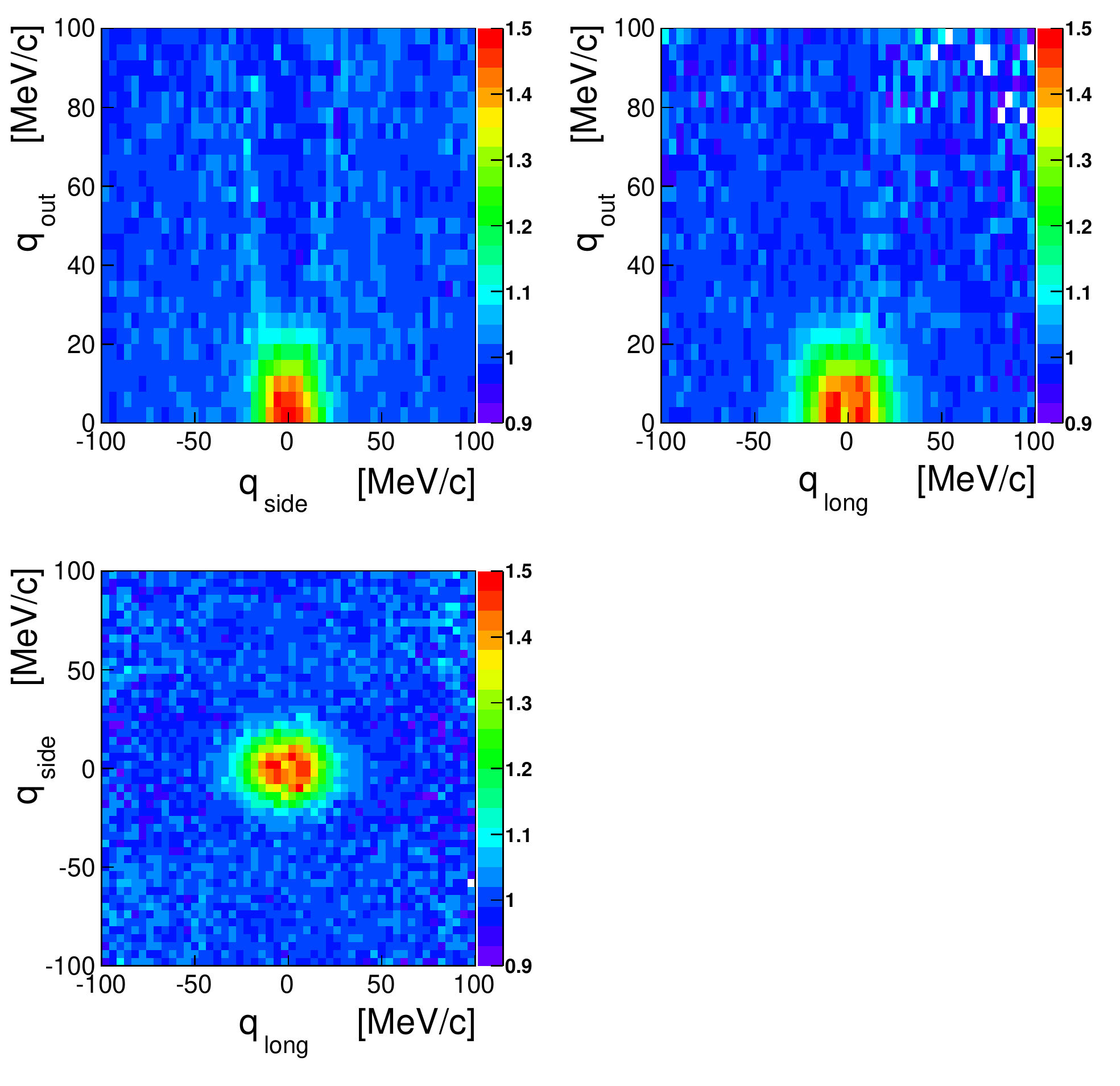}
\end{center}
\caption{
Two-dimensional projections of the Coulomb-corrected 
three-dimensional correlation function (Eq.\,(\ref{pipi_fit_fct_3dim})) 
in the 'out' vs. 'side' (upper left panel), 'out' vs. 'long' 
(upper right), and 'side' vs. 'long' (lower left) planes, respectively, 
for $\pi^-\pi^-$ pairs with transverse momentum of  $p_\mathrm{t,\,12}=100-400~\mevc$ 
and angles w.r.t. the reaction plane of $\vert\phi_{12}-\phi_{\mathrm{RP}}\vert<\pi/8$ 
for centralities of $10-30\,\%$. The respective third direction is integrated over 
$\pm 20~\mevc$. }
\label{fig:Cosl_centr10to30_pt100to400_phi0}
\end{figure}
\begin{figure}[h]
\begin{center}
\includegraphics[width=0.9\linewidth]{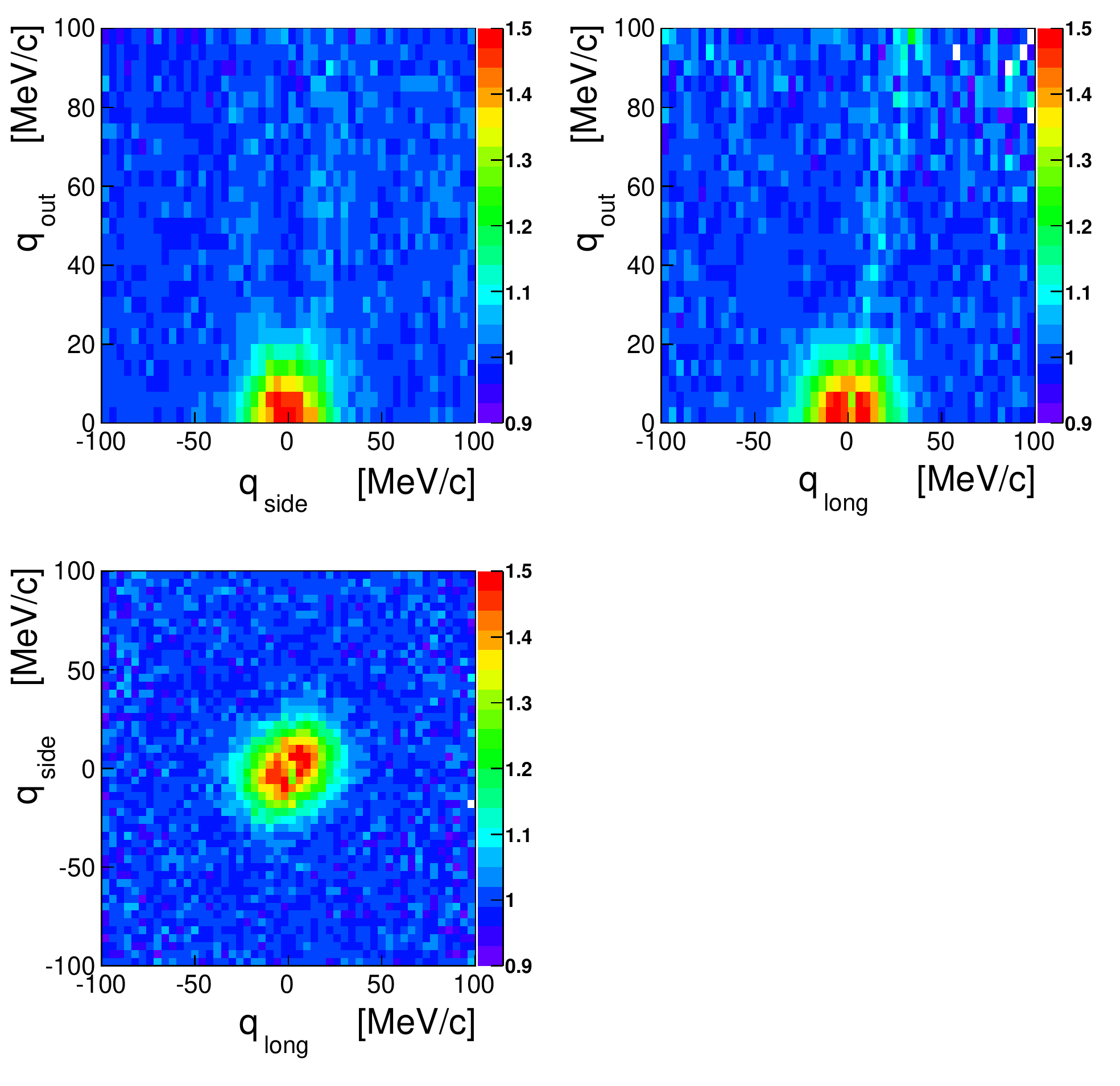}
\end{center}
\caption{The same as Fig.\,\ref{fig:Cosl_centr10to30_pt100to400_phi0}, 
but for  $\vert\phi_{12}-\phi_{\mathrm{RP}}-\pi/2\vert<\pi/8$. }
\label{fig:Cosl_centr10to30_pt100to400_phi90}
\end{figure}
\subsection{Separation of central charge bias - construction of neutral pion radii}
\label{sect:pi0pi0_corr}
To quantify a potential source radius bias introduced by the Coulomb force the charged pions 
experience in the field of the charged fireball, we follow the ansatz used in 
ref.\,\cite{Baym_1996,Baym_1997}, 
\begin{equation}
E({\bm p}_\mathrm{f}) = E({\bm p}_\mathrm{i}) \pm V_\mathrm{eff}({\bm r}_\mathrm{i}),
\label{eq:baym_1}
\end{equation}
where ${\bm p}_\mathrm{i}$ (${\bm p}_\mathrm{f}$) is the initial (final) momentum, $E$ the 
corresponding total energy, and ${\bm r}_\mathrm{i}$ the inital position of the pion  
in the Coulomb potential $V_{\mathrm{eff}}$ with positive (negative) sign for $\pi^+$ ($\pi^-$). 
With
\begin{align}
 \frac{R_{\pi^{\pm}\pi^{\pm}}}{R_{\pi^0\pi^0}} \approx \frac{q_\mathrm{i}}{q_\mathrm{f}} &= \frac{|{\bm p}_\mathrm{i}|}{|{\bm p}_\mathrm{f}|} \nonumber \\ 
                                               &= \sqrt{ 1 \mp 2 \frac{V_{\mathrm{eff}}}{|{\bm p}_\mathrm{f}|} \sqrt{1 + \frac{m_{\pi}^2}{{\bm p}_\mathrm{f}^2}} + \frac{V^2_{\mathrm{eff}}}{{\bm p}_\mathrm{f}^2}}, 
\end{align}
where $q_\mathrm{i}$ ($q_\mathrm{f})$ is the initial (final) relative momentum, and with  
$V_{\mathrm{eff}} / k_\mathrm{t} \ll 1$, it turns out that the squared source radius for 
pairs of constructed neutral pions 
(denoted by $\tilde{\pi}^0\tilde{\pi}^0$ in the following)
is simply the arithmetic mean of the corresponding quantities of the charged pions,  
\begin{equation}
R_{\tilde\pi^0\tilde\pi^0}^2 = \frac{1}{2} \big(R_{\pi^+\pi^+}^2 + R_{\pi^-\pi^-}^2\big).
\label{eq:baym_2}
\end{equation}
Finally, the constructed $\tilde{\pi}^{0}\tilde{\pi}^{0}$ correlation radii, discussed in the 
subsequent sections, are derived from cubic spline interpolations of the $k_\mathrm{t}$ and 
$A_\mathrm{part}$ dependences of the corresponding experimental $\pi^{-}\pi^{-}$ and 
$\pi^{+}\pi^{+}$ data. 

\subsection{Azimuthally-integrated HBT analysis}
\label{sect:ana_azi_int}
\subsubsection{Central collisions}
\label{sect:radii_central}
We start with the study of source radii for central ($0-10\,\%$) collisions. 
Figure\,\ref{fig:Radien_pi0pi0extract_spline_centr0to10_pt0to900} shows the $m_\mathrm{t}$ 
dependence of the one-dimensional (invariant) and three-dimensional source radii for $\pi^-\pi^-$ 
(black squares) and $\pi^+\pi^+$ (red circles) pairs. While for low transverse mass the 
Coulomb interaction with the fireball leads to an increase (a decrease) of the source size 
derived for negative (positive) pion pairs, at large transverse momentum the 
Coulomb effect apparently fades away. The effect is smallest for $\rout$ 
and most pronounced for $\rside$. Note that the charge 
splitting of the source radii 
as well as its difference in magnitude in out and side direction
was early on predicted by Barz~\cite{Barz96,Barz99} 
who investigated the combined effects of nuclear Coulomb field, radial flow, and opaqueness on 
two-pion correlations for a large collision system such as Au\,+\,Au in the $1 \agev$ energy 
regime. Earlier experimental works at the Bevalac employing a three-body Coulomb correction 
found the effect negligible for their studies of smaller systems 
\cite{Zajc84,Christie92,Christie93}. We note that the extent of the source 
in direction of ${\bm k}_\mathrm{t}$ is potentially enlarged by a finite emission 
time duration, which is expected to last a few fm\,/$c$ \cite{Lisa05}, 
cf. Sect.\,\ref{sect:ana_azi_sens}. 
It would be very interesting to study the charge-sign effect
quantitatively with the help of dynamical models of the collision evolution. At present, 
such theoretical investigations concentrate on the higher energies where the charge splitting 
is hardly visible.

The parameter $\losl$ derived from the fits with Eq.\,(\ref{pipi_fit_fct_3dim}) appears 
to be rather independent of transverse mass and charge sign and decreases only slightly with 
increasing transverse mass, cf. lower right panel of 
Fig.\,\ref{fig:Radien_pi0pi0extract_spline_centr0to10_pt0to900}. It 
fits well into a preliminary evolution with $\sqrt{s_\mathrm{NN}}$ established 
previously \cite{STAR2015}, except the lowest E895 data point at 2\agev. In contrast, $\linv$ 
resulting from the fits to the one-dimensional correlation function, 
exhibits a significant decrease with $m_\mathrm{t}$ (cf. lower left panel), 
probably pointing to the fact that the one-dimensional fit function is not adequate.   
\begin{figure}[h]
\begin{center}
\includegraphics[width=0.9\linewidth]{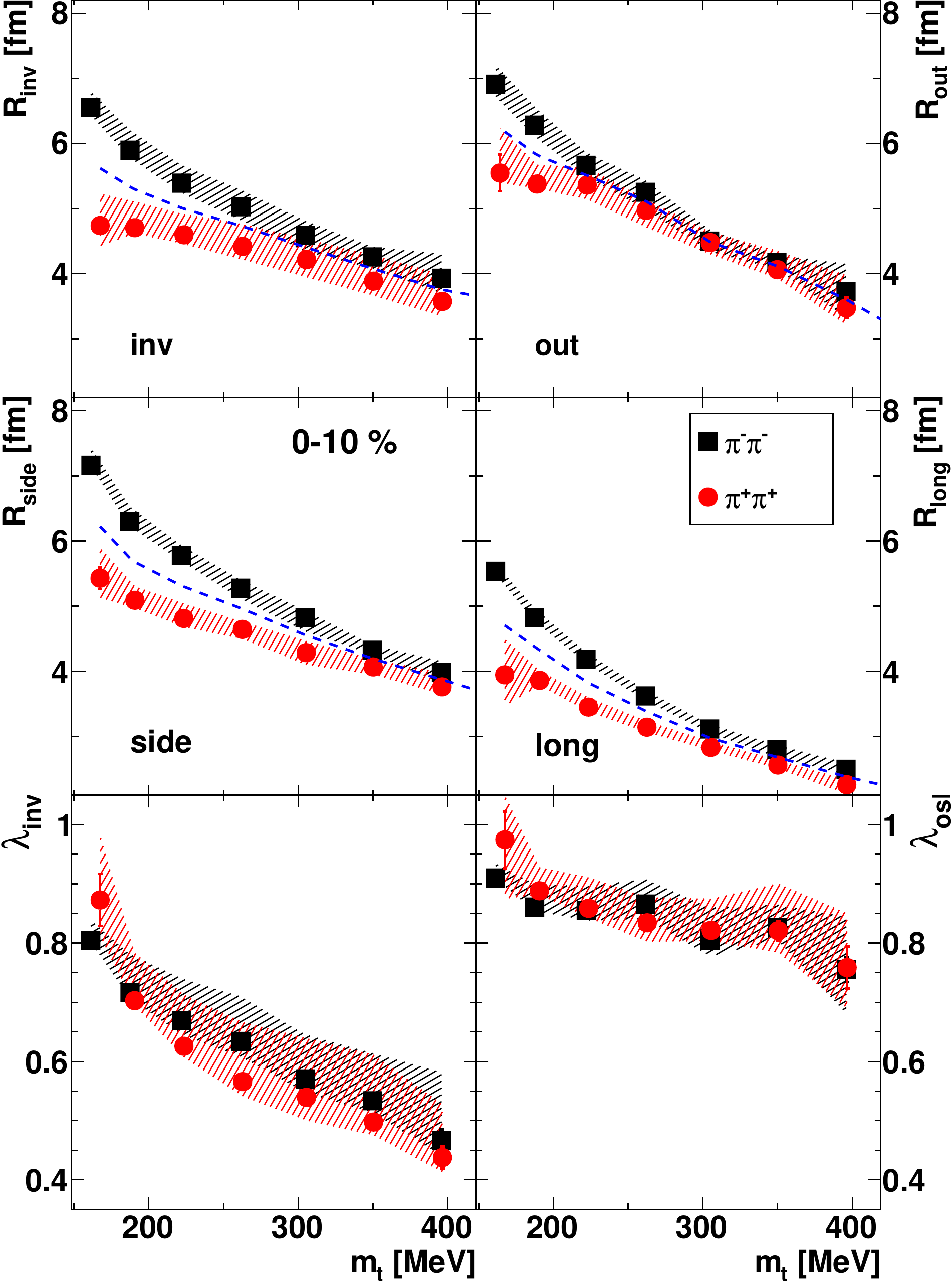}
\end{center}
\caption{Source radii as function of pair transverse mass, $m_\mathrm{t}$, for central 
($0-10\,\%$) \AuAu. The corresponding $p_\mathrm{t,\,12}$ range amounts to $100-800~\mevc$.  
The upper left, upper right, center left, and center right panels display the invariant, 'out', 
'side', and 'long' radii, respectively.  The lower left and lower right panels show the 
corresponding $\lambda$ parameters resulting from the fits to the one- and three-dimensional 
correlation functions, respectively. 
Black squares (red circles) are for pairs of negative 
(positive) pions. Blue dashed curves represent constructed radii (Eq.\,(\ref{eq:baym_2}))
of neutral pion pairs. Error bars and hatched bands  
represent the statistical and systematic errors, respectively.  }
\label{fig:Radien_pi0pi0extract_spline_centr0to10_pt0to900}
\end{figure}
\subsubsection{Centrality dependence}
\label{sect:centrality}
Figure\,\ref{fig:Radien_inv_osl_pimpim_Apartdepend_pt300to400} exhibits the $A_\mathrm{part}$ 
dependence of source radii of $\pimin\pimin$ pairs (upper left panel) 
for the transverse-momentum interval $p_\mathrm{t,\,12} = 300-400~\mevc$. 
The volume of the overlap region (i.e. the participant zone) of the colliding nuclei is 
expected to be proportional to $A_{\mathrm{part}}$. Consequently, the involved length scales 
should be proportional to $A_{\mathrm{part}}^{1/3}$, which therefore could provide a good 
scaling variable for emission source radii. 
Close to linear dependences of the pion source radii 
as function of $A_{\mathrm{part}}^{1/3}$ are observed, 
as demonstrated by the straight-line fits to the data. 
The upper right panel of Fig.\,\ref{fig:Radien_inv_osl_pimpim_Apartdepend_pt300to400}
shows the same for $\piplus\piplus$ source radii,  
following similar linear dependences on $A_{\mathrm{part}}^{1/3}$. 
Note that the $\rinv(A_\mathrm{part})$ dependences (black dotted 
lines) of the charged pion pairs, when extrapolated down to the $A_\mathrm{part}$ value 
(not shown) of the 
system Ar\,+\,KCl previously investigated by HADES \cite{hades_hbt_ArKCl},   
match the corresponding radii within uncertainties. This observation 
supports the idea that the variation of the participant volume via both, 
the choice of the size of the collision partners and the selection of a certain collision  
centrality, is equivalent. 
Following the recipe given in Sect.\,\ref{sect:pi0pi0_corr} to remove the 
Coulomb effect from the above dependences of $\pi^{-}\pi^{-}$ and $\pi^{+}\pi^{+}$ radii,   
we present the $A_\mathrm{part}^{1/3}$ dependence of constructed 
$\tilde{\pi}^{0}\tilde{\pi}^{0}$ source radii 
in the lower left panel of Fig.\,\ref{fig:Radien_inv_osl_pimpim_Apartdepend_pt300to400}. 

All results of the azimuthally integrated one-dimensional 
$(\rinv,\,\linv)$ and three-dimensional $(\rout,\,\rside,\,\rlong,\,\losl)$ fits in 
dependence on collisions centrality and mean transverse momentum  
for $\pi^-\pi^-$, $\pi^+\pi^+$ and constructed $\tilde{\pi}^{0}\tilde{\pi}^{0}$ pairs 
are summarized in tables\,\ref{tab:pimpim}, \ref{tab:pippip} and \ref{tab:pi0pi0},  
respectively.

Finally, the transverse-mass dependences of constructed $\tilde{\pi}^{0}\tilde{\pi}^{0}$
radii for different centrality classes are summarized 
in Fig.\,\ref{fig:Radien_inv_osl_pi0pi0_allcentralities_newbinning_pt100to800}. 
The data are fitted (dashed curves) with a function 
\begin{align}
R= R_{0} \left(\frac{m_\mathrm{t}}{m_\pi}\right)^{\alpha}.
\label{fit_R0_alpha}
\end{align} 
The fit parameters, $R_0=R(k_\mathrm{t}=0)$ and $\alpha$, derived from the $\chi^2$ 
minimizations are summarized in table\,\ref{tab:alpha_R0}. 
Here we emphasize that, while $\alpha$ derived from a fit to 
$\rside(m_\mathrm{t})$ of $\tilde\pi^0\tilde\pi^0$ follows the expectation (i.e. $\alpha=-0.5$)
of (3+1)D hydrodynamics \cite{Kisiel2014},  
its absolute value is larger (smaller) by $\sim 0.1$ when  
fitting the corresponding dependences of $\pi^{-}\pi^{-}$ ($\pi^{+}\pi^{+}$).   
\begin{figure}[h]
\begin{center}
\includegraphics[width=0.9\linewidth]{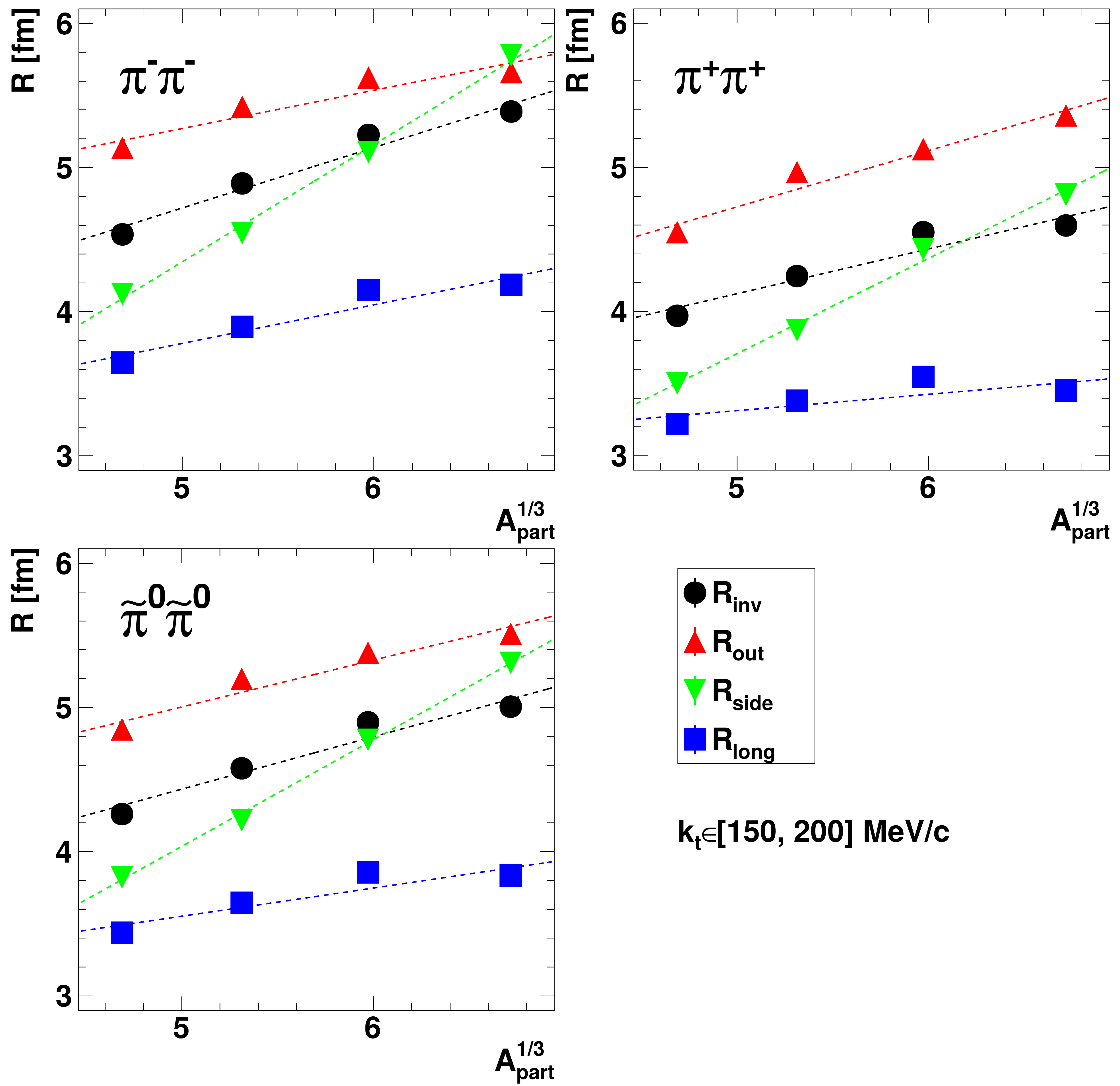}
\end{center}
\caption{Source radii as function of the cube root of the number of participants 
for $\pimin\pimin$ (upper left) and $\piplus\piplus$ (upper right) pairs with transverse 
momenta of $p_\mathrm{t,\,12}=300-400~\mevc$.  The lower left panel 
shows corresponding radii of constructed neutral pions as deduced from 
interpolation of the charged-pion data according to the procedure decribed in 
Sect.\,\ref{sect:pi0pi0_corr}.   
Black circles, red triangles, green inverted triangles, and blue squares 
display the 'invariant', 'out', 'side', and 'long' radii, respectively. Dotted lines represent  
$\chi^2$ fits to the data with straight lines.  }
\label{fig:Radien_inv_osl_pimpim_Apartdepend_pt300to400}
\end{figure}
\begin{figure}[h]
\begin{center}
\includegraphics[width=0.9\linewidth]{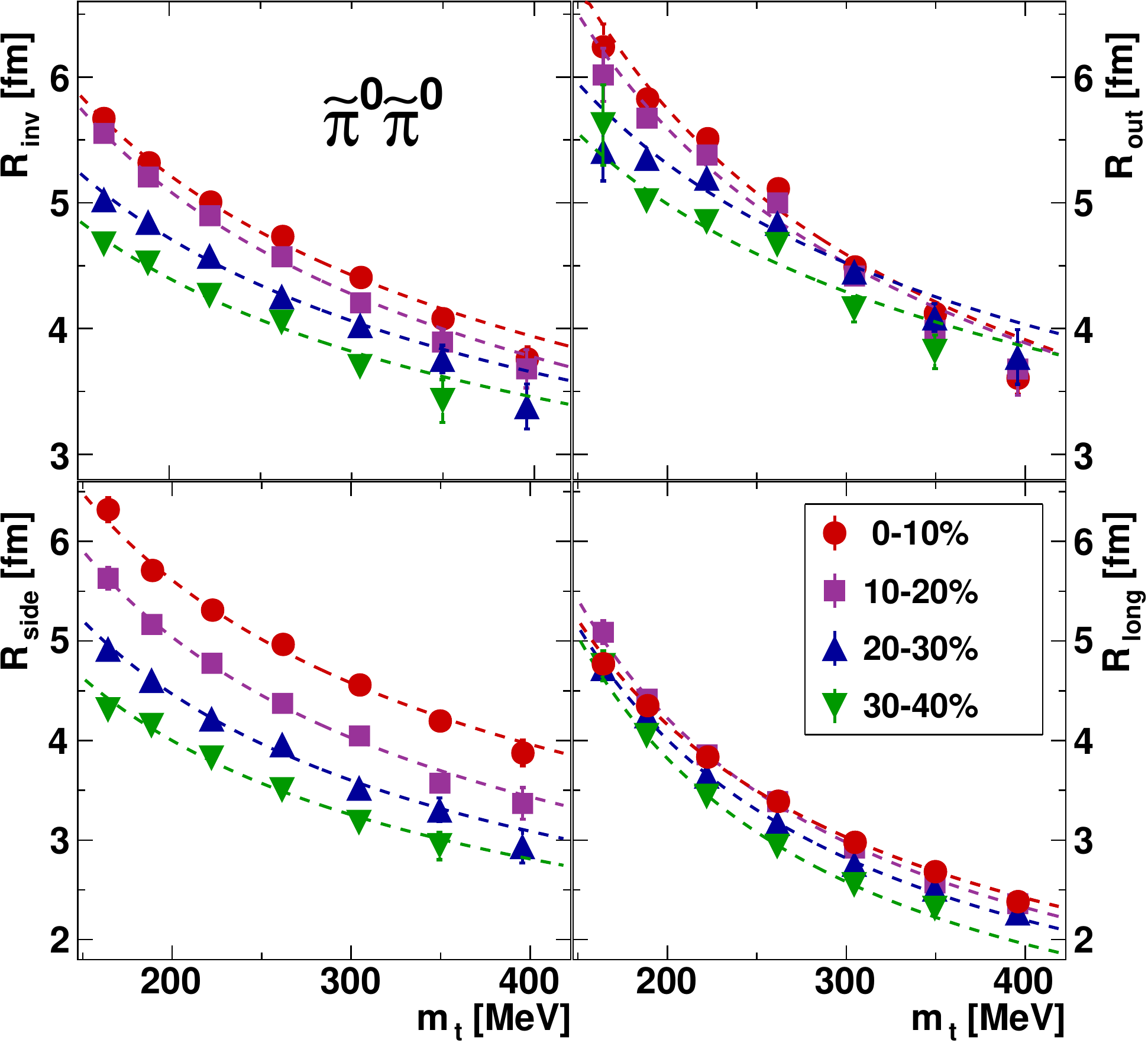}
\end{center}
\caption{
Source radii as function of the transverse mass for pairs of constructed neutral 
pions as deduced from interpolations of the charged-pion data according to the procedure of 
Sect.\,\ref{sect:pi0pi0_corr}. The upper left, upper right, lower left, and lower right panels 
display the 'invariant', 'out', 'side', and 'long' radii, respectively. Red circles, magenta  
squares, blue triangles, and green inverted triangles represent the centrality 
classes $0-10\,\%$, $10-20\,\%$, $20-30\,\%$, and $30-40\,\%$, 
respectively. Dashed curves are $\chi^2$ fits to the data with a function   
$R= R_0 (m_\mathrm{t}/m_\pi)^\alpha$. The fit parameters $R_0$ and $\alpha$ are summarized in 
table\,\ref{tab:alpha_R0}.  }
\label{fig:Radien_inv_osl_pi0pi0_allcentralities_newbinning_pt100to800}
\end{figure}
\subsection{Azimuthally-sensitive HBT analysis}
\label{sect:ana_azi_sens}
To estimate the geometrical quantities hidden in the azimuthal variation of the 
correlation function (Eq.\,(\ref{pipi_fit_fct_3dim})), we follow the recipe given bei 
Wiedemann and Heinz \cite{Wiedemann_1998,Wiedemann_Heinz_1999} 
and perform a common fit to our data on squared radii from  
Eq.\,(\ref{pipi_fit_be_3dim_general}), using the entire set of fit equations (Eqs.\,(2)) 
given in ref.\,\cite{e895_2000_phi}, which yields the elements of the spatial correlation 
tensor as ten angle-independent fit parameters:  
\begin{align}
S_{\mu\nu}=\langle\tilde x_{\mu}\,\tilde x_{\nu}\rangle,\,\,\tilde x_{\mu}=x_{\mu}-\langle x_{\mu} \rangle,\hskip 10bp (\mu,\nu=0,\,1,\,2,\,3).
\label{S_mu_nu}
\end{align} 
Here, $x_0=t$ is the time component, $x_1=x$ is parallel to 
the impact parameter $\bm{b}$, and $x_3=z$ points in beam direction. The Cartesian coordinate 
system is completed with direction $x_2=y$ being perpendicular to the reaction plane formed by 
$x$ and $z$. The brackets $\langle\,\rangle$ indicate an average over the emission source. 

A typical fit result is  
displayed in Fig.\,\ref{fig:pimpim_azimuthal_Rosl_centr10to30_pt300to400}  
for $\pimin\pimin$ pairs with average transverse momentum of $\bar k_\mathrm{t}=170~\mevc$ 
and for centralities of $10-30\,\%$. It delivers the uncorrected matrix $S$  
(in units of fm$^2$, with statistical errors attached):    
\begin{align}
& S^{\mathrm{\,meas}} =  \nonumber \\ 
& \begin{pmatrix}~10.12 \pm 0.64 & -0.77 \pm 0.21 &  -0.15 \pm 0.23 &  ~~~0.09\pm 0.14  \\
                  -0.77 \pm 0.21 & ~21.60 \pm 0.28 &  -0.07 \pm 0.19 &  ~~~2.10\pm 0.11  \\
                  -0.15 \pm 0.23 & -0.07 \pm 0.19 &  ~28.87 \pm 0.34 &  -0.05\pm 0.11  \\
                 ~~~0.09 \pm 0.14 & ~~~2.10 \pm 0.11 &  -0.05 \pm 0.11 &  ~16.12\pm 0.13 
\end{pmatrix}.  
\label{S_meas}
\end{align}
As expected from symmetry considerations \cite{Lisa_Heinz_Wiedemann_2000}, 
only the diagonal elements and $S_{13}$ differ  
significantly from zero, yielding the following six non-vanishing squared radii
\begin{align}
R^2_{\mathrm{out}} =\,&\,\frac{1}{2}(S_{11}+S_{22})+\frac{1}{2}(S_{22}-S_{11})\cos{(2\Phi)}+\langle\beta^2_t\rangle S_{00}, \nonumber 
\\
R^2_{\mathrm{side}}=\,&\,\frac{1}{2}(S_{11}+S_{22})+\frac{1}{2}(S_{22}-S_{11})\cos{(2\Phi)}, \nonumber  
\\
R^2_{\mathrm{long}}=\,&\,S_{33}+\langle\beta^2_l\rangle S_{00}, \nonumber
\\
R^2_{\mathrm{outside}}=\,&\,\frac{1}{2}(S_{22}-S_{11})\sin{(2\Phi)}, \nonumber
\\
R^2_{\mathrm{outlong}}=\,&\,S_{13}\cos{(\Phi)}, \nonumber
\\
R^2_{\mathrm{sidelong}}=\,&\,-S_{13}\sin{(\Phi)},
\label{R2_azi} 
\end{align}
where $\beta_\mathrm{t}$ and $\beta_\mathrm{l}$ are the pair velocities 
in transverse and longitudinal direction, respectively. 

The correction of both, the finite reaction-plane resolution and the finite azimuthal bin width,  
is performed following the method described in refs. \cite{Ollitrault97,HBT_in_UrQMD}:  
\begin{align}
R^{2,\mathrm{corr}}_{\mathrm{i,n}} = R^{2,\mathrm{meas}}_{\mathrm{i,n}} \, \dfrac{n \, \Delta/2}{F_{\mathrm{n}} \sin(n \, \Delta/2)}, 
\label{RP_resol}
\end{align}
where $R^{2,\mathrm{corr}}_{\mathrm{i,n}}$ 
($i=$ 'out', 'side', 'long', 'outside', 'outlong', 'sidelong',\, $n=0,\,1,\,2$)
are the underlying (``true'') Fourier coefficients \cite{HBT_in_UrQMD},  
$\Delta=\pi/4$ is the present $\Phi$ interval, and the quantity $F_{\mathrm{n}}$ 
represents the $n$-th event-plane resolution\footnote{$S_{01}$ $S_{02}$, $S_{13}$, 
and $S_{23}$ undergo a $n=1$ correction.   
$S_{12}$ and $S_{22}-S_{11}$ are subjected to a $n=2$ correction. 
$S_{00}$, $S_{11}+S_{22}$, $S_{33}$, and $S_{03}$ are not affected by Eq.\,(\ref{RP_resol}).}. 
The values of $F_1$ and 
$F_2$ for the centrality classes investigated in the present analysis 
are summarized in table\,\ref{tab:rp_resolution}. For the detailed description of the 
event-plane reconstruction and the centrality dependence of  the resolution parameters 
$F_\mathrm{n}$ we refer to a forthcoming HADES paper on 
the $n$-th ($n \le 4$) collective flow observables of protons and light nuclei produced 
in \AuAu~\cite{hades_flow_AuAu}.  
\begin{figure}[h]
\begin{center}
\includegraphics[width=0.9\linewidth]{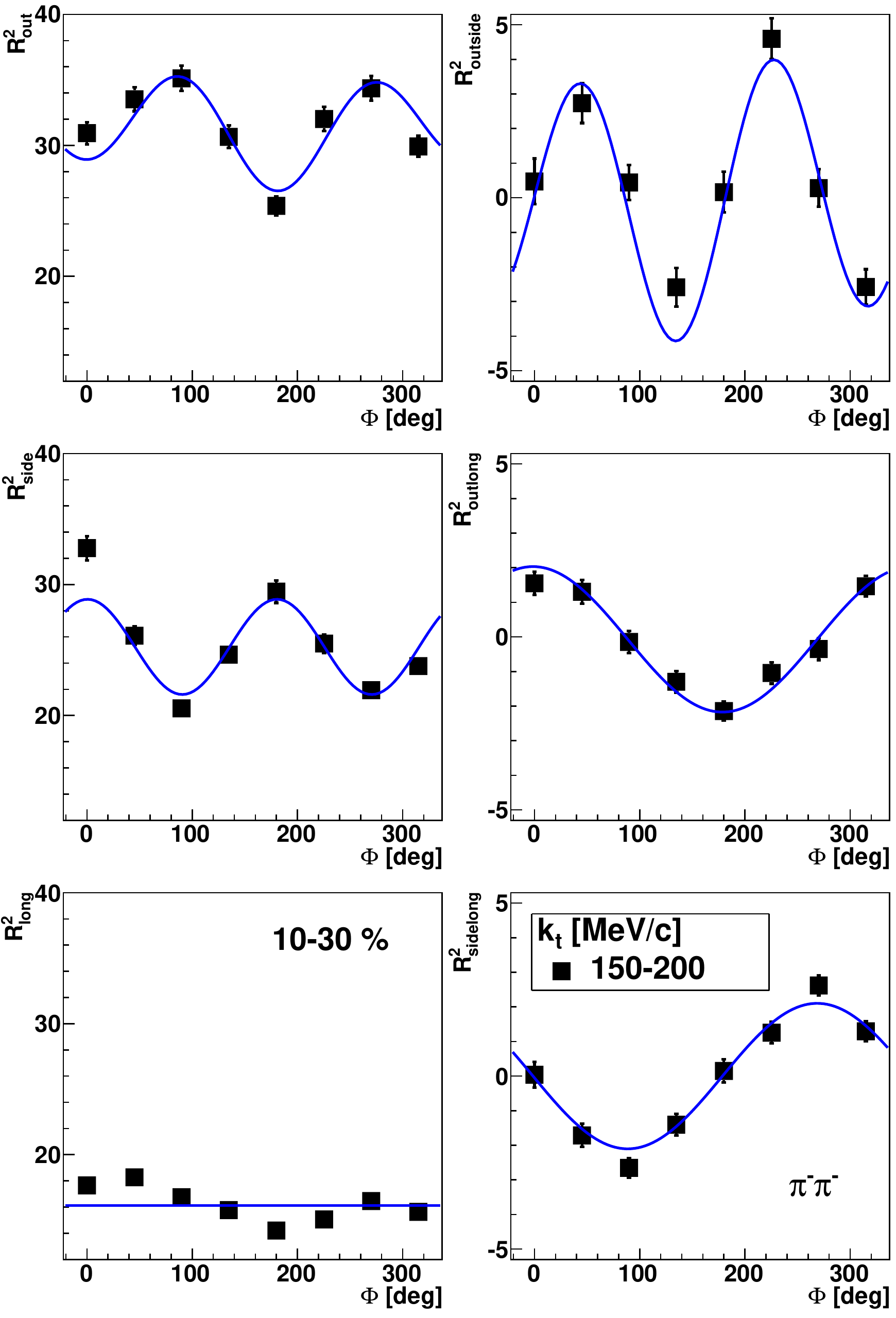}
\end{center}
\caption{Squared radii as function of the pair azimuthal angle relative to the reaction plane, 
$\Phi=\phi_{12}-\phi_\mathrm{RP}$, for $\pi^-\pi^-$ pairs with transverse momenta of 
$p_\mathrm{t,\,12}=300-400~\mevc$ and for centralities of $10-30\,\%$. 
The left column from top to bottom  
shows the 'out', 'side', and 'long' radii. The right column gives the 'outside', 
'outlong', and 'sidelong' components. Error bars include only statistical uncertainties. 
The full curves represent a global fit with Eqs.\,(2) of ref.\,\cite{e895_2000_phi} to the 
experimental data points. }
\label{fig:pimpim_azimuthal_Rosl_centr10to30_pt300to400}
\end{figure}

The matrix $S$ (Eq.\,(\ref{S_meas})) after the correction (Eq.\,(\ref{RP_resol})) reads 
\begin{align}
& S^{\mathrm{\,corr}} =  \nonumber \\ 
& \begin{pmatrix} ~10.12\pm 0.64 &  -0.91\pm 0.25 &  -0.18\pm 0.27 &   ~~~0.09\pm 0.14 \\
                  -0.91\pm 0.25 &  ~18.60\pm 0.40 &  -0.13\pm 0.36 &   ~~~2.49\pm 0.13  \\
                  -0.18\pm 0.27 &  -0.13\pm 0.36 &  ~31.87\pm 0.44 &  -0.06\pm 0.14  \\
                   ~~~0.09\pm 0.14 &   ~~~2.49\pm 0.13 &  -0.06\pm 0.14 &  ~16.12\pm 0.13 
\end{pmatrix}.
\label{S_corr} 
\end{align}
From this matrix the spatial tilt angle in the reaction plane can be calculated as 
\begin{align}
\theta_\mathrm{s}=\frac{1}{2}\tan^{-1}\left(\frac{2 S^\mathrm{\,corr}_{13}}{S^\mathrm{\,corr}_{33}-S^\mathrm{\,corr}_{11}}\right)=(-32\pm2)^{\circ}.
\label{tilt}
\end{align}
Rotating $S^{\mathrm{\,corr}}$ by the angle $-\theta_\mathrm{s}$ around the $y$ axis, 
i.e. applying the corresponding rotation matrix $G_y(\theta_\mathrm{s})$, yields a diagonal tensor  
\begin{align}
& S^{\mathrm{\,diag}} = G^{\dag}_y(\theta_\mathrm{s})\cdot S^{\mathrm{\,corr}}\cdot G_y(\theta_\mathrm{s}) = \nonumber \\ 
& \begin{pmatrix} ~10.12\pm 0.64 &  -0.73\pm 0.23 &  -0.18\pm 0.27 &   ~~~0.56\pm 0.18 \\
                  -0.73\pm 0.23 &  ~20.15\pm 0.35 &  -0.15\pm 0.31 &   ~~~0.00\pm 0.23 \\
                  -0.18\pm 0.27 &  -0.15\pm 0.31 &  ~31.87\pm 0.44 &   ~~~0.02\pm 0.22 \\
                   ~~~0.56\pm 0.18 &  ~~~0.00\pm 0.23 &   ~~~0.02\pm 0.22 &  ~14.58\pm 0.24 
\end{pmatrix}
\label{S_diag} 
\end{align}
whose eigenvalues are the temporal and geometrical variances 
$\sigma^2_\mathrm{t}, \sigma^2_\mathrm{x}, \sigma^2_\mathrm{y}, \sigma^2_\mathrm{z}$.  
The geometrical variances of $\pimin\pimin$ emission as function of transverse-momentum  
are displayed in the left column of Fig.\,\ref{fig:pimpim_azi_sig_tilt_ecc} 
for all centralities. Figure\,\ref{fig:pimpim_azi_sig_tilt_ecc} also shows the 
transverse-momentum dependences of the tilt angle, $\theta_\mathrm{s}$, the $xy$-eccentricity, 
\begin{align}
\epsilon_{xy}=\frac{\sigma_\mathrm{y}^2-\sigma_\mathrm{x}^2}{\sigma_\mathrm{x}^2+\sigma_\mathrm{y}^2},
\label{eps_xy}
\end{align}
and the $zy$-eccentricity,  
\begin{align}
\epsilon_{zy}=\frac{\sigma_\mathrm{y}^2-\sigma_\mathrm{z}^2}{\sigma_\mathrm{y}^2+\sigma_\mathrm{z}^2}.
\label{eps_zy}
\end{align}
Note that we included an intermediate centrality class of  $25-35\,\%$ to 
verify the strong centrality dependence of the tilt angle 
(upper right panel of Fig.\,\ref{fig:pimpim_azi_sig_tilt_ecc} and discussion below). 
For all transverse momenta and all centralities, the 
deduced eccentricities represent an almond shape in the plane perpendicular to the beam 
direction. The  $xy$-eccentricity becomes smallest (almost circular shape) for most 
central collisions. Almost no centrality dependence appears for the $zy$-eccentricity. 
The range of variation with transverse momentum is larger for the $zy$-eccentricity as 
compared to the one in the $xy$-plane. For large momenta, $\theta_\mathrm{s}$ tends to 
vanish, provided that the 30\,\% most central event classes      
(i.e. impact parameters not considerably larger than 50\,\% of the maximum one) are selected,  
and the final $\pimin\pimin$ $xy$-eccentricity recovers the corresponding initial 
eccentricity. No charge-sign difference appears in the transverse-momentum 
dependence of both, the tilt angle and the eccentricities, while the spatial principal axes 
differ, as demonstrated in Fig.\,\ref{fig:pimpim_pippip_azi_sig_tilt_ecc_centr10to30} for  
medium centralities of $10-30\,\%$. 

Due to the freedom in the coordinate-system definition, the tilt angle, defined as the angle 
between the $z$ coordinate (directed along the shortest principal axis in our case) and the 
beam direction, can be changed from $\theta_\mathrm{s}$ to $\theta_\mathrm{s}-90^o$, while 
$\sigma_x$ and $\sigma_z$ (and accordingly $\epsilon_{xy}$ and 
$\epsilon_{xy}$) interchange. 
The arrangement of our data (cf. Fig.\,\ref{fig:pimpim_azi_sig_tilt_ecc}) was choosen such that 
both dependences, on transverse momentum and on centrality, show smooth trends. Thus, 
smaller $\vert\theta_\mathrm{s}\vert$ for more central collisions are ensured, as one would 
expect from collision geometry. However, at more peripheral collisions and high 
values of $p_{\mathrm{t},12}$ a different configuration is conceivably, i.e. requiring 
$\theta_\mathrm{s} = 0$ at high transverse momenta (relevant here only for the data 
points at $p_{\mathrm{t},12} = 440 \, \mathrm{MeV}/c$ and centrality $25-35\,\%$). Within our 
statistical and systematic uncertainties we are not able to find a final decision, which 
arrangement is the better one. (Note also that, near $p_{\mathrm{t},12} = 400 \, \mathrm{MeV}/c$ 
and for centralities beyond $20\,\%$, $\sigma_x$ and $\sigma_z$ are about the same size, 
which disturbes the picture of a well defined tilt angle, eventually represented by large 
uncertainties of this quantity.)

The volume of the region of homogeneity, 
\begin{align} 
V_\mathrm{fo}=(2\pi)^{3/2}\sigma_{\mathrm{x}} \sigma_{\mathrm{y}} \sigma_{\mathrm{z}},
\label{Vfo_3sigma}
\end{align} 
as derived from the spatial principal axes of the Gaussian emission ellipsoid is presented 
in Fig.\,\ref{fig:Vfo_vs_pt_pimpim_pippip_centr_sep} (full symbols) as function of 
transverse momentum for all centrality classes. For comparison, also the results for the 
approximate volume, 
\begin{align}
V_\mathrm{fo}'=(2\pi)^{3/2}R^2_{\mathrm{side}}R_{\mathrm{long}},
\label{Vfo_2RsideRlong} 
\end{align}
following from the azimuthally-integrated analysis (open symbols) is shown. 
Differences are significant only at small transverse momenta, 
where the tilt angle vanishes. 

To study the centrality dependence of the tilt angle in more detail, 
Fig.\,\ref{fig:pimpim_pippip_azi_tilt_b} displays $\theta_\mathrm{s}$ for $\pimin\pimin$ 
(full symbols) and $\piplus\piplus$ (open symbols) pairs as function of the impact parameter, 
$b$ (cf. table\,\ref{tab:rp_resolution}), for different average transverse momentum values.  
While for lower momenta the magnitude of $\theta_\mathrm{s}$ is proportional to $b$, 
this dependence gets weaker with increasing pair transverse momentum until the tilt is  
close to zero for the highest momentum classes.
No charge-sign difference is observed in the centrality dependence of $\theta_\mathrm{s}$. 

Figure\,\ref{fig:pimpim_pippip_azi_ecc_ini_fin} explicitly  
relates the $xy$-eccentricity for $\pi^-\pi^-$ pairs to the initial eccentricity 
relative to the participant plane (cf. Eq.\,(21) of ref.\,\cite{STAR2015}),  
as derived from Glauber simulations \cite{PhobosGlauber2015}. 
For large transverse momenta the source eccentricity derived from the present 
identical-pion HBT analysis recovers the inital (nucleon) eccentricity. 
This observation is essentially different from the picture at higher energies,
where the $k_\mathrm{t}$ integrated freeze-out eccentricity is found considerably below the 
corresponding initial one \cite{STAR2015,PHENIX2014,ALICE2017}. The trend of 
increasing freeze-out eccentricity with increasing transverse momentum is also found by ALICE 
at LHC, though on a much lower absolute scale \cite{ALICE2017}. The model of     
(3+1)D hydrodynamics \cite{Kisiel2014} slightly unterestimates the final source 
eccentricity at LHC. It would be very desirable to discuss in separate future works 
the present observations at SIS18 energies in the framework of contemporary hydro or 
transport models. 

Fig.\,\ref{fig:pimpim_azi_time}  
shows the temporal components of the spatial correlation tensor (cf. Eqs.\,(2) of 
ref.\,\cite{e895_2000_phi}). For the emission duration, $\sigma_\mathrm{t}^2=S_{00}$, the fit  
yields values significantly deviating from zero, 
especially for low transverse momenta and for all considered 
centrality classes except the most central one. 
The values for the mixed elements, however, are much  
smaller and are mostly consistent with zero, with the possible exception of pion pairs with 
small transverse momenta in central collisions. As already mentioned above, the disappearance 
of $S_{01}$, $S_{02}$ and $S_{03}$ is expected due to symmetry reasons  
\cite{Lisa_Heinz_Wiedemann_2000}. However, one should keep in mind that the symmetry of  
the rapidity distribution w.r.t. midrapidity is not perfect in 
the present fixed-target experiment, even though the rapidity interval was rather 
restricted for this analysis ($\vert y - y_{cm} \vert < 0.35$).

Finally, the geometrical and temporal variances and the tilt angle 
of the $\pimin\pimin$ ($\piplus\piplus$) emission source 
in dependence of centrality and transverse momentum 
are summarized in table\,\ref{tab:pimpim_azimuthal} (\ref{tab:pippip_azimuthal}). 
\begin{figure}[h]
\begin{center}
\includegraphics[width=0.9\linewidth]{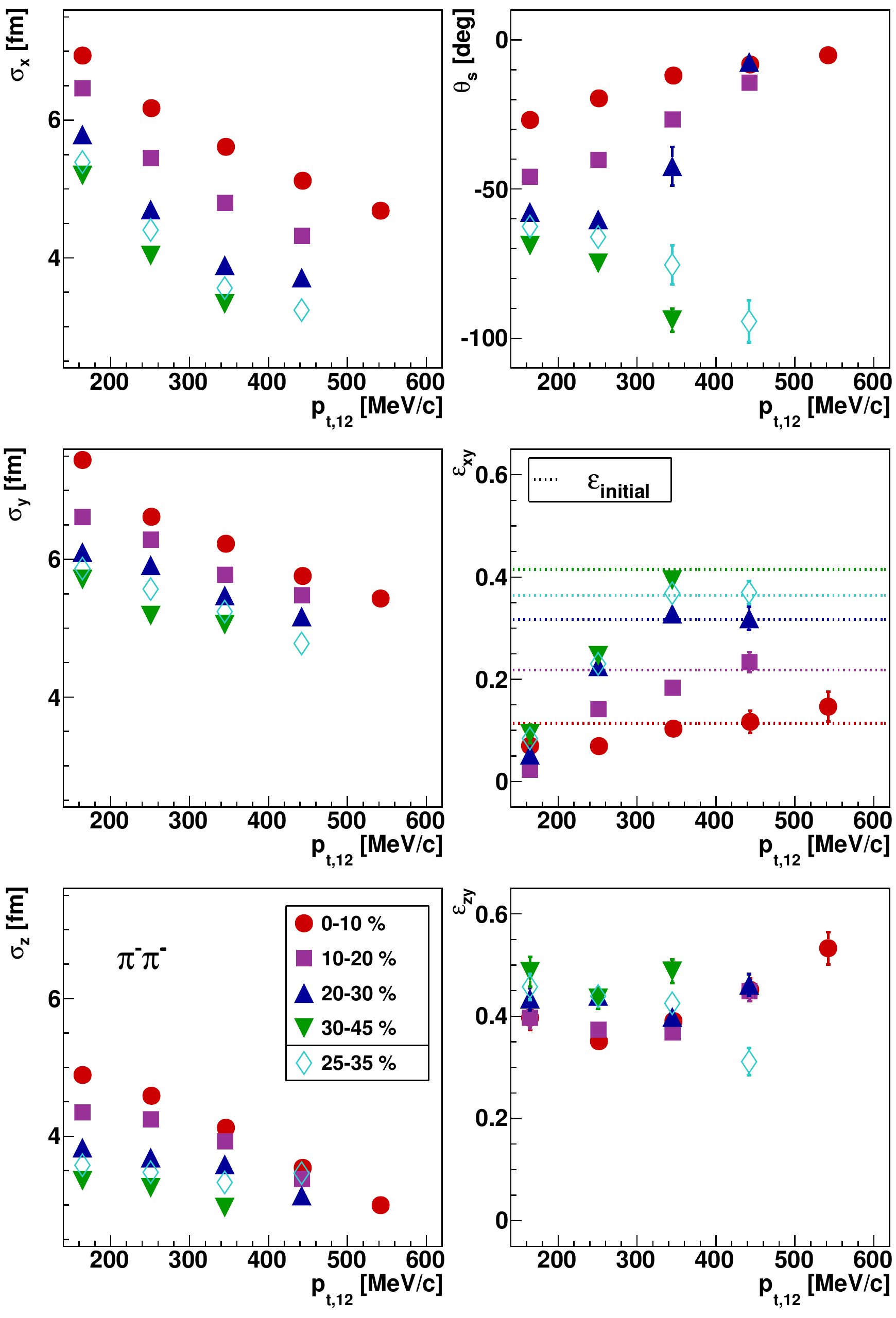}
\end{center}
\caption{The spatial principal axes (left column),the tilt angle w.r.t. the beam axis in the 
reaction plane (Eq.\,(\ref{tilt}), top right), the $xy$-eccentricity  
(Eq.\,(\ref{eps_xy}, center right), and the $zy$-eccentricity 
(Eq.\,(\ref{eps_zy}, bottom right) of the Gaussian emission ellipsoid of $\pimin\pimin$ 
pairs as function of pair transverse momentum for different centrality classes (cf. legend). 
Error bars include only statistical uncertainties. Dotted lines  
represent the corresponding initial eccentricity as derived from Glauber simulations. } 
\label{fig:pimpim_azi_sig_tilt_ecc}
\end{figure}
\begin{figure}[h]
\begin{center}
\includegraphics[width=0.9\linewidth]{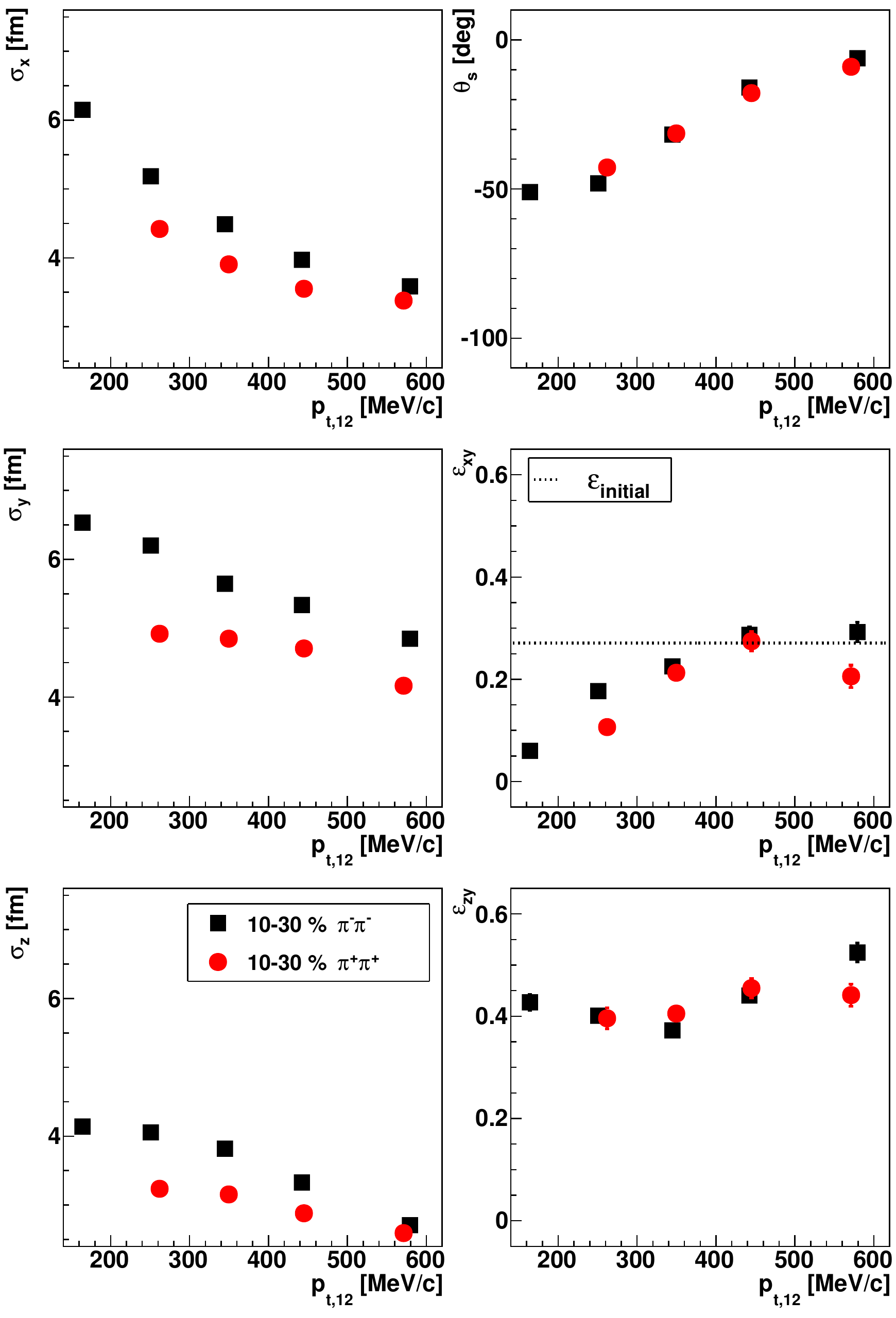}
\end{center}
\caption{The same as Fig.\,\ref{fig:pimpim_azi_sig_tilt_ecc}, but for $\pimin\pimin$ 
(black squares) and $\piplus\piplus$ (red circles)
pairs in comparison for medium centralities of $10-30\,\%$. } 
\label{fig:pimpim_pippip_azi_sig_tilt_ecc_centr10to30}
\end{figure}
\begin{figure}[h]
\begin{center}
\includegraphics[width=0.9\linewidth]{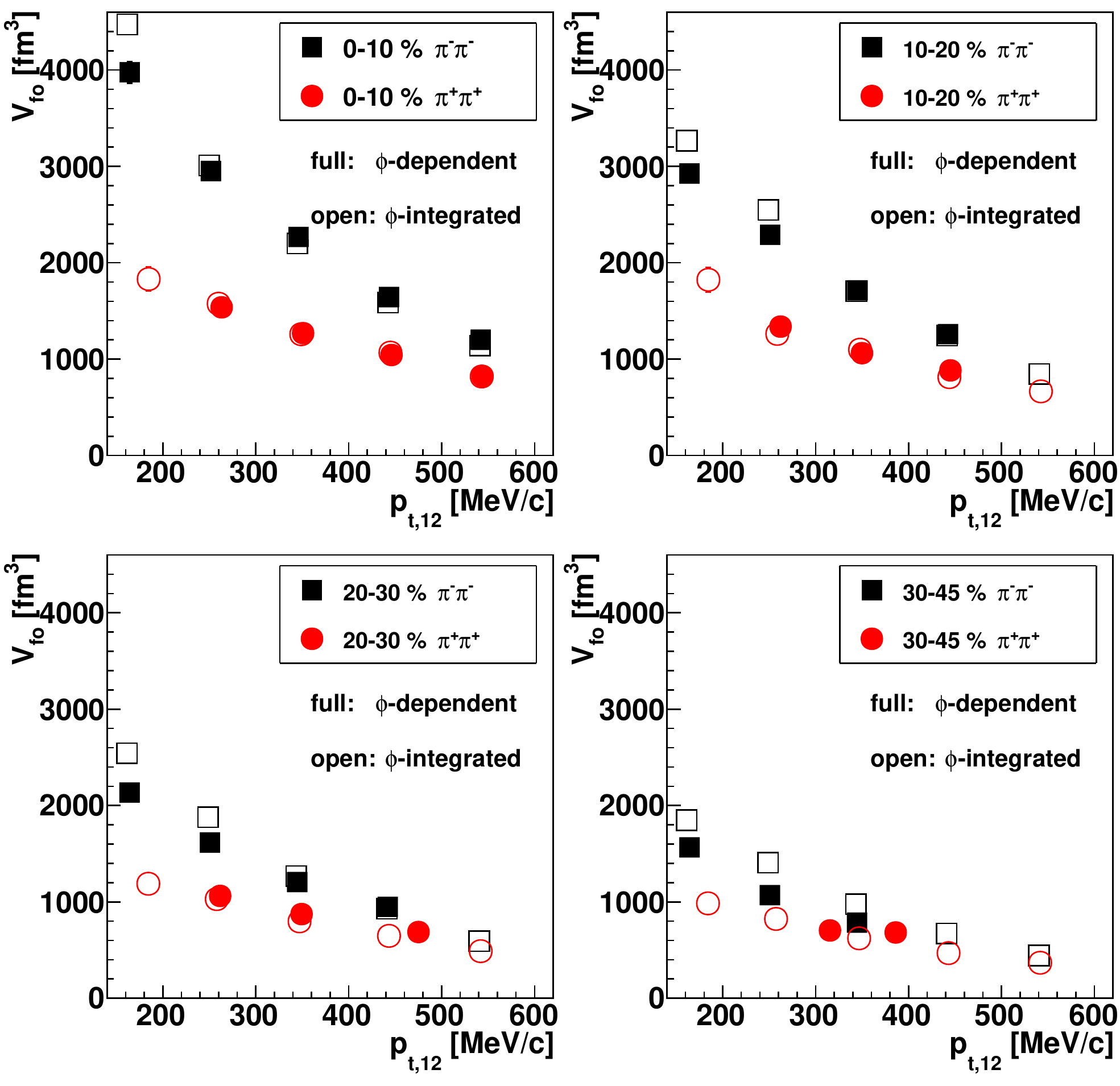}
\end{center}
\caption{The volume of the region of homogeneity 
(Eq.\,(\ref{Vfo_3sigma}), full symbols) as derived from the 
azimuthally-sensitive analysis and the approximate volume 
(Eq.\,(\ref{Vfo_2RsideRlong}), open symbols) following from the azimuthally-integrated analysis  
of $\pimin\pimin$ (squares) and $\piplus\piplus$ (circles) pairs as function of pair transverse 
momentum for different centralities (cf. legend). } 
\label{fig:Vfo_vs_pt_pimpim_pippip_centr_sep}
\end{figure}
\begin{figure}[h]
\begin{center}
\includegraphics[width=0.9\linewidth]{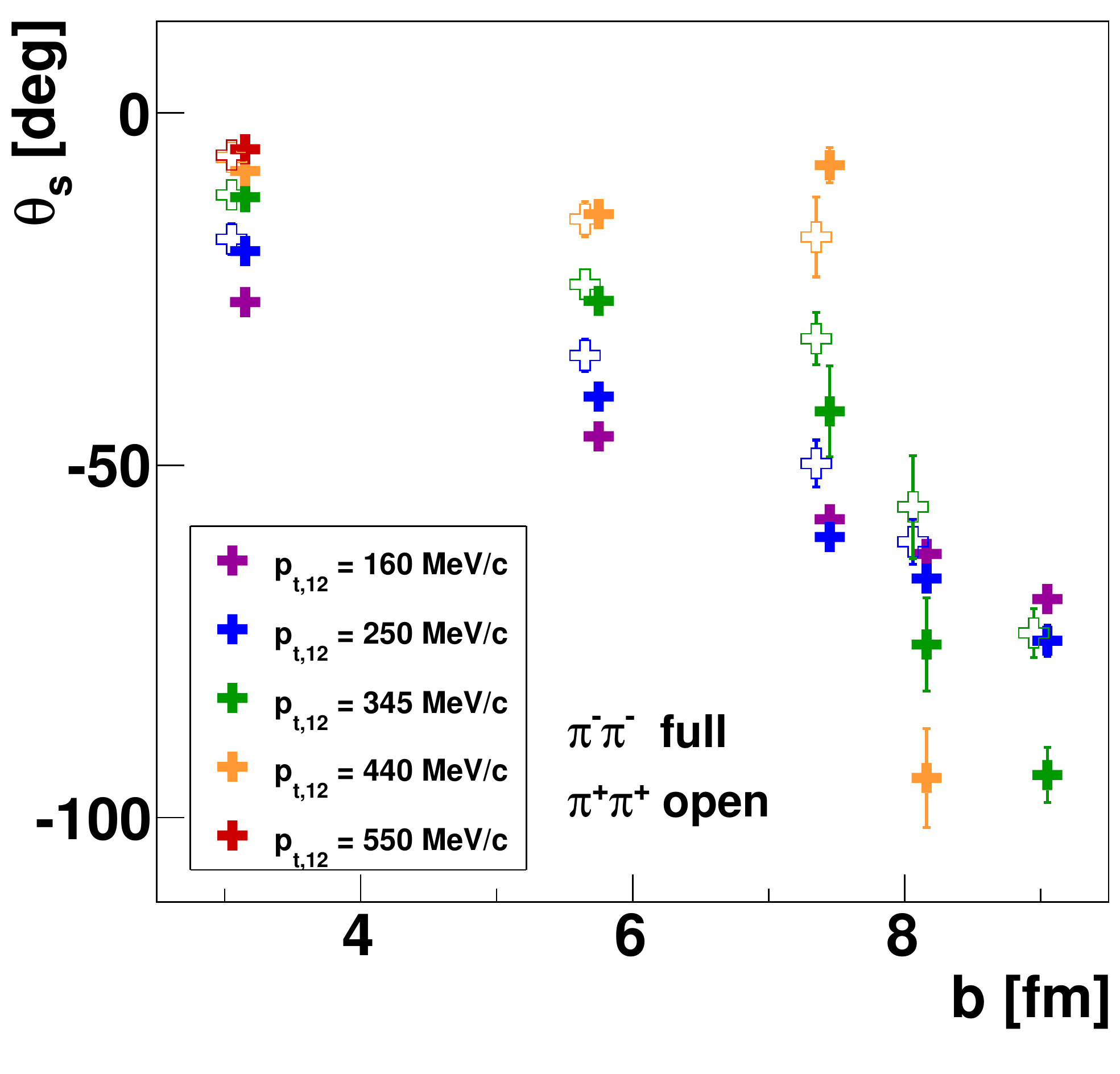}
\end{center}
\caption{The emission-ellipsoid tilt angle, $\theta_\mathrm{s}$, in the reaction plane for  
$\pimin\pimin$ (full symbols) and $\piplus\piplus$ (open symbols) pairs as function of the 
mean impact parameter derived from Glauber MC calculations \cite{hades_centrality:2018am} 
for different pair transverse momentum classes. Error bars include only statistical uncertainties.}
\label{fig:pimpim_pippip_azi_tilt_b}
\end{figure}
\begin{figure}[h]
\begin{center}
\includegraphics[width=0.9\linewidth]{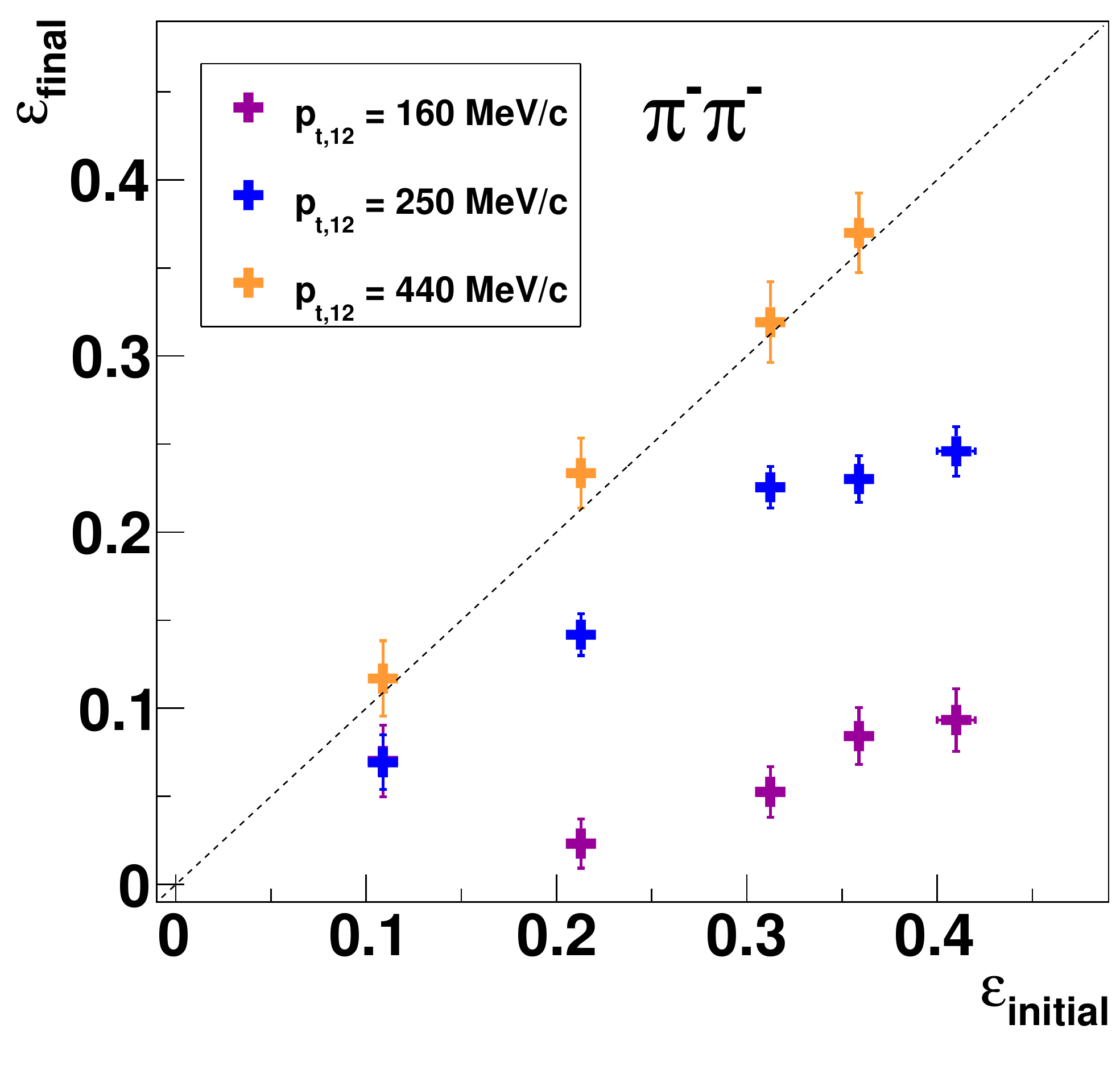}
\end{center}
\caption{The $\pimin\pimin$ emission ellipsoid $xy$-eccentricity (Eq.\,(\ref{eps_xy})) as function 
of the initial nucleon eccentricity derived from Glauber simulations \cite{PhobosGlauber2015} 
for different pair transverse momentum classes. 
Error bars include only statistical uncertainties. The dashed line indicates 
$\epsilon_{\mathrm{final}} = \epsilon_{\mathrm{initial}}$. }
\label{fig:pimpim_pippip_azi_ecc_ini_fin}
\end{figure}
\begin{figure}[h]
\begin{center}
\includegraphics[width=0.9\linewidth]{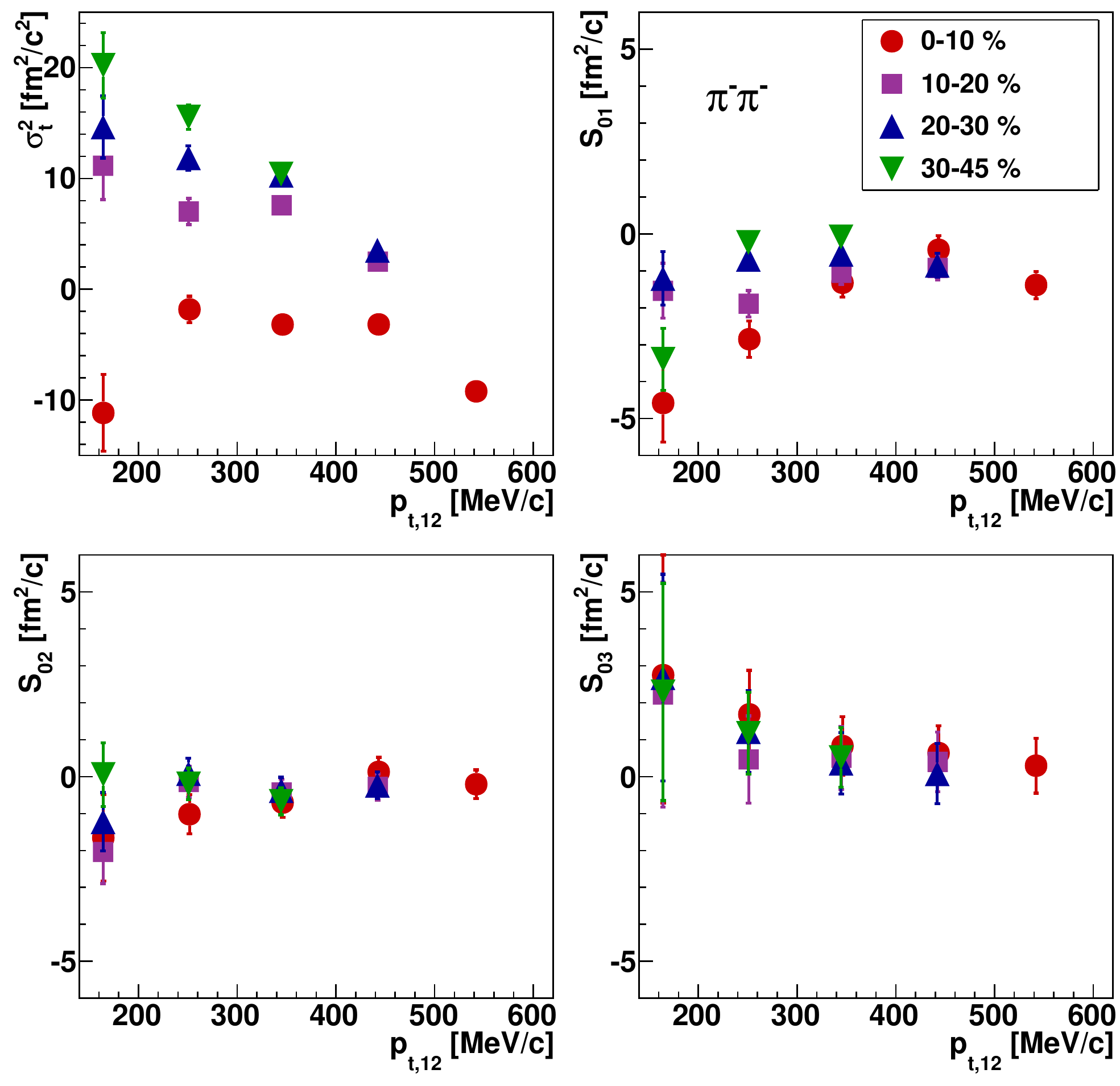}
\end{center}
\caption{The temporal components of the spatial correlation tensor $S$ (Eq.\,(\ref{S_mu_nu})) 
for $\pimin\pimin$ pairs as function of transverse momentum for different centrality 
classes (cf. legend). The upper left, upper right, lower left, and lower right panels display 
$\sigma_\mathrm{t}^2=S_{00}$, $S_{01}$, $S_{02}$, and  $S_{03}$, respectively. Error bars 
include only statistical uncertainties. }
\label{fig:pimpim_azi_time}
\end{figure}
\subsection{Comparison with other experiments - Excitation functions of source parameters}
\label{sect:exc_func}
The energy dependences below $\sqrt{s_{\mathrm{NN}}}=4$~GeV 
of the $\pimin\pimin$ emission ellipsoid principal axes, 
the tilt angle w.r.t. the beam axis in the reaction plane, the eccentricities, 
and the emission duration for medium centralities 
and average transverse momentum of $\bar k_\mathrm{t}=110~\mevc$ are shown in 
Fig.\,\ref{fig:pimpim_azi_sig_tilt_ecc_excitationfct}.
The displayed HADES/SIS18 data follow the trends of the E895/AGS data  
\cite{e895_2000_phi}. 

The excitation function of eccentricities over a wider range of collision energies 
is presented in Fig.\,\ref{fig:pipi_ecc_excfct}. While for the HADES and E895 data 
the exact equations (\ref{eps_xy}) and (\ref{eps_zy}) are applied 
(cp. Fig.\,\ref{fig:pimpim_azi_sig_tilt_ecc_excitationfct}), for $\sqrt{s_{\mathrm{NN}}}>4$~GeV, 
where no $\pi\pi$ emission source data after principal-axis transformation 
of the spatial correlation tensor are available, the approximations 
$\epsilon_\mathrm{xy}\approx 2R^2_{\mathrm{side,\,2}}/R^2_{\mathrm{side,\,0}}$ and 
$\epsilon_{zy}\approx 1- 2R^2_{\mathrm{long,\,0}}/(R^2_{\mathrm{side,\,0}}+2R^2_{\mathrm{side,\,2}}+R^2_{\mathrm{long,\,0}})$ 
are used. Here, the quantities $R^2_{\mathrm{side,\,0}}$ 
($R^2_{\mathrm{long,\,0}}$) and $R^2_{\mathrm{side,\,2}}$ are the zeroth- and 2nd-order Fourier 
coefficients of the azimuthal-angle parameterization of the 'side' ('long') 
radius, respectively, where $2 R^2_{\mathrm{side,\,2}}$ 
corresponds to the oscillation amplitude of both the squared 'out' and 'side' radii  
(cf. Eqs.\,(\ref{R2_azi}) and Fig.\,\ref{fig:pimpim_azimuthal_Rosl_centr10to30_pt300to400}). 
These approximations are justified by the small tilt angles found at high collision energies. 
We observe a remarkable increase of both types of eccentricity at low $\sqrt{s_\mathrm{NN}}$. 
It would be very desirable to study the strong increase towards low energy of both 
eccentricity and tilt angle in the framework of contemporary models. First investigations of 
both observables by means of several theoretical transport calculations qualitatively reproduced  
the experimental findings \cite{Lisa2011}; the quantitative reproduction of their dependences on 
collision energy, centrality and transverse momentum should allow for further 
discrimination among the models. 

The excitation function of the squared emission duration which is either taken directly from 
the fit parameter $S_{00}$, i.e. the temporal variance of the $\pi\pi$ emission 
(cf. Eqs.\,(\ref{R2_azi})), or from the difference of the squared 'out' and 'side' radii, 
$(\Delta\tau)^2=(R_\mathrm{out}^2-R_\mathrm{side}^2)/\langle \beta^2_\mathrm{t}\rangle$, 
is shown in Fig.\,\ref{fig:dtau_exc_fct}. For all data of the various experiments, 
the emission duration appears small, i.e. typically a few fm/$c$. 
With increasing transverse momentum, the emission duration deduced from the 
HADES data is found to decrease, virtually vanishing for large momenta. 
Taking the data points at $k_\mathrm{t}=310~\mevc$, there is a clear energy dependence: 
A rise towards $\sqrt{s_\mathrm{NN}}\sim 10-20$~GeV and then a a slow drop towards LHC.   

Finally, we return to a few results of the azimuthally-integrated HBT analysis, concentrating 
on central collisions ($0-10\,\%$). 
The excitation functions of $R_{\mathrm{out}}$, $R_{\mathrm{side}}$, and $R_{\mathrm{long}}$ 
for pion pairs produced in most central events are displayed 
in Fig.\,\ref{fig:Rosl_exctfct_mt260_central_pimpim_pi0pi0}. All shown radius parameters have been 
obtained by interpolating the existing measured data points to the same 
transverse mass of $m_\mathrm{t}=330~\mev$ ($k_\mathrm{t}=300~\mevc$)
at which the charge differences of the source radii 
almost vanish (cf. Fig.\,\ref{fig:Radien_pi0pi0extract_spline_centr0to10_pt0to900}). 
The statistical errors are properly propagated and 
quadratically added to the differences between linear and cubic-spline interpolations. 
Extrapolations were not necessary at this $m_\mathrm{t}$ value. $R_{\mathrm{out}}$ 
and $R_{\mathrm{side}}$ vary not more than 40\,\% over three orders of magnitude in 
center-of-mass energy. Only $R_{\mathrm{long}}$ exhibits a steady 
increase by about a factor of two when going in energy from SIS18 via AGS, SPS, RHIC  
to LHC. Note that in the excitation functions shown in ref.\,\cite{STAR2015} not all, 
particularly AGS, data points were properly corrected for their $k_\mathrm{t}$ dependence. 

The excitation function of $R_{\mathrm{out}}^2-R_{\mathrm{side}}^2$ for an average  
transverse momentum of the pion pairs of $k_\mathrm{t}=300~\mevc$ in central collisions 
is shown in Fig.\,\ref{fig:Rout2_Rside2_mt260_central_pimpim_pi0pi0}. 
Almost all other measurements below 10\,GeV are characterized by large errors and scatter 
sizeably. The new HADES data show that the difference of the source parameters in the 
transverse plane almost vanishes at low collision energies. Since this quantity is related to the 
emission duration via Eqs.\,(\ref{R2_azi}), one would conclude that    
in the 1\agev~energy region the observed pions are emitted into free space during a short 
time span of less than a few fm/$c$ (cp. also 
Figs.\,\ref{fig:pimpim_azi_sig_tilt_ecc_excitationfct}  
(bottom right) and \ref{fig:dtau_exc_fct} displaying similar data divided by 
$\langle \beta^2_\mathrm{t}\rangle$ for centralities of $0-10\,\%$ and $10-30\,\%$, respectively, 
but for different transverse momenta). 
However, also the opaqueness of the source affects 
$R_{\mathrm{out}}^2-R_{\mathrm{side}}^2$ which could cause it to become negative, thus 
compensating the positive contribution of the emission duration \cite{Barz99}. 
We also emphasize that, with increasing available 
energy, this quantity reaches a local maximum at $\sqrt{s_{\mathrm{NN}}}\sim 20-30$\,GeV and 
afterwards decreases towards zero at LHC energies. 
 
The excitation function of the approximate volume of the region of 
homogeneity, Eq.\,(\ref{Vfo_2RsideRlong}),  
for central collisions is given in Fig.\,\ref{fig:Vfreezeout_mt160_central_pimpim_pi0pi0}. 
Here, we chose this approximation, in contrast to  Eq.\,(\ref{Vfo_3sigma}), for the 
sake of comparability with other experiments. Note that this definition of a three-dimensional 
Gaussian volume does not incorporate $\rout$ since this length 
is potentially extended due to a finite value of the aforementioned emission duration. 
For the large transverse momenta selected, the differences of the 
azimuthally-integrated and azimuthally-sensitive analyses as well as the charge-sign 
splitting largely vanish, cp. Fig.\,\ref{fig:Vfo_vs_pt_pimpim_pippip_centr_sep} exhibiting 
the strong transverse-momentum dependence of the volume, especially for $\pimin\pimin$ pairs.  
From the above HADES data, we estimate a volume of about 850\,fm$^3$ for 
constructed $\tilde{\pi}^0\tilde{\pi}^0$ pairs. This volume of homogeneity 
steadily increases with energy, but appears merely a factor four larger at LHC. Extrapolating 
the volume to $k_\mathrm{t}=0$ yields a value of about 3,900\,fm$^3$. 

Similar excitation functions at other transverse momenta or centralities (with proper 
interpolation/extrapolation of the transverse-momentum and centrality dependences) 
can be derived from the source parameters summarized in tables \ref{tab:pimpim}, \ref{tab:pippip}, 
\ref{tab:pimpim_azimuthal}, and \ref{tab:pippip_azimuthal}.
\begin{figure}[h]
\begin{center}
\includegraphics[width=0.9\linewidth]{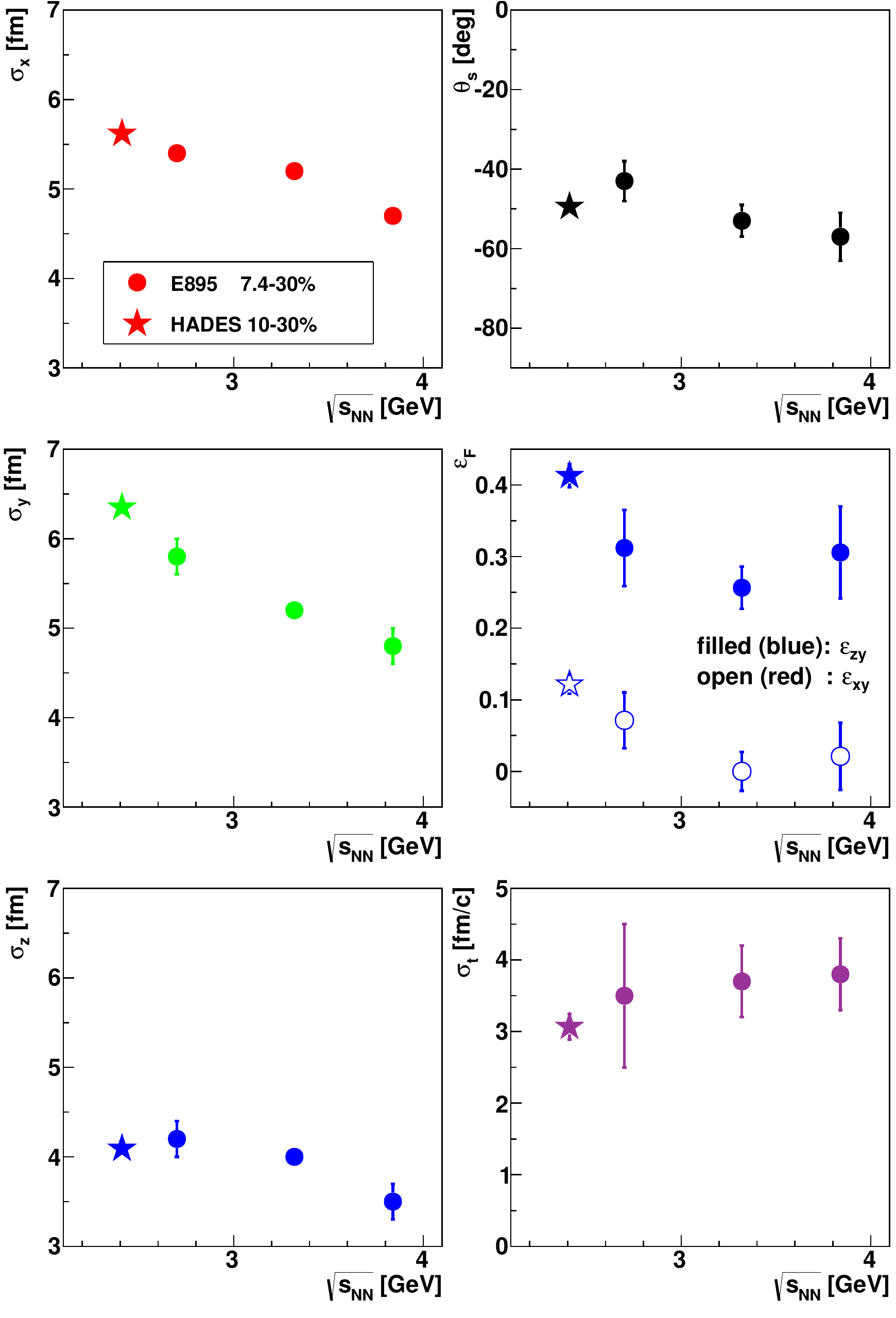}
\end{center}
\caption{The low-energy excitation function of the principal axes (left column), the tilt angle 
in the reaction plane (Eq.\,(\ref{tilt}), top right), the eccentricities $\epsilon_{xy}$ 
(Eq.\,(\ref{eps_xy}), center right, open symbols) and $\epsilon_{zy}$ (Eq.\,(\ref{eps_zy}), 
center right, filled symbols), and the emission duration $\sigma_\mathrm{t}=\sqrt{S_{00}}$ 
(Eqs.\,(\ref{R2_azi}), bottom right) as derived from the Gaussian emission ellipsoid of 
$\pimin\pimin$ pairs. Stars are HADES data for centralities of $10-30\,\%$ 
and average transverse momentum of $\bar k_\mathrm{t}=110~\mevc$. 
The circles are corresponding E895 data for slightly different centralities taken at 
beam energies of 2, 4, and 6~\agev~\cite{e895_2000_phi}. 
Error bars include only statistical uncertainties.  }
\label{fig:pimpim_azi_sig_tilt_ecc_excitationfct}
\end{figure}
\begin{figure}[h]
\begin{center}
\includegraphics[width=0.9\linewidth]{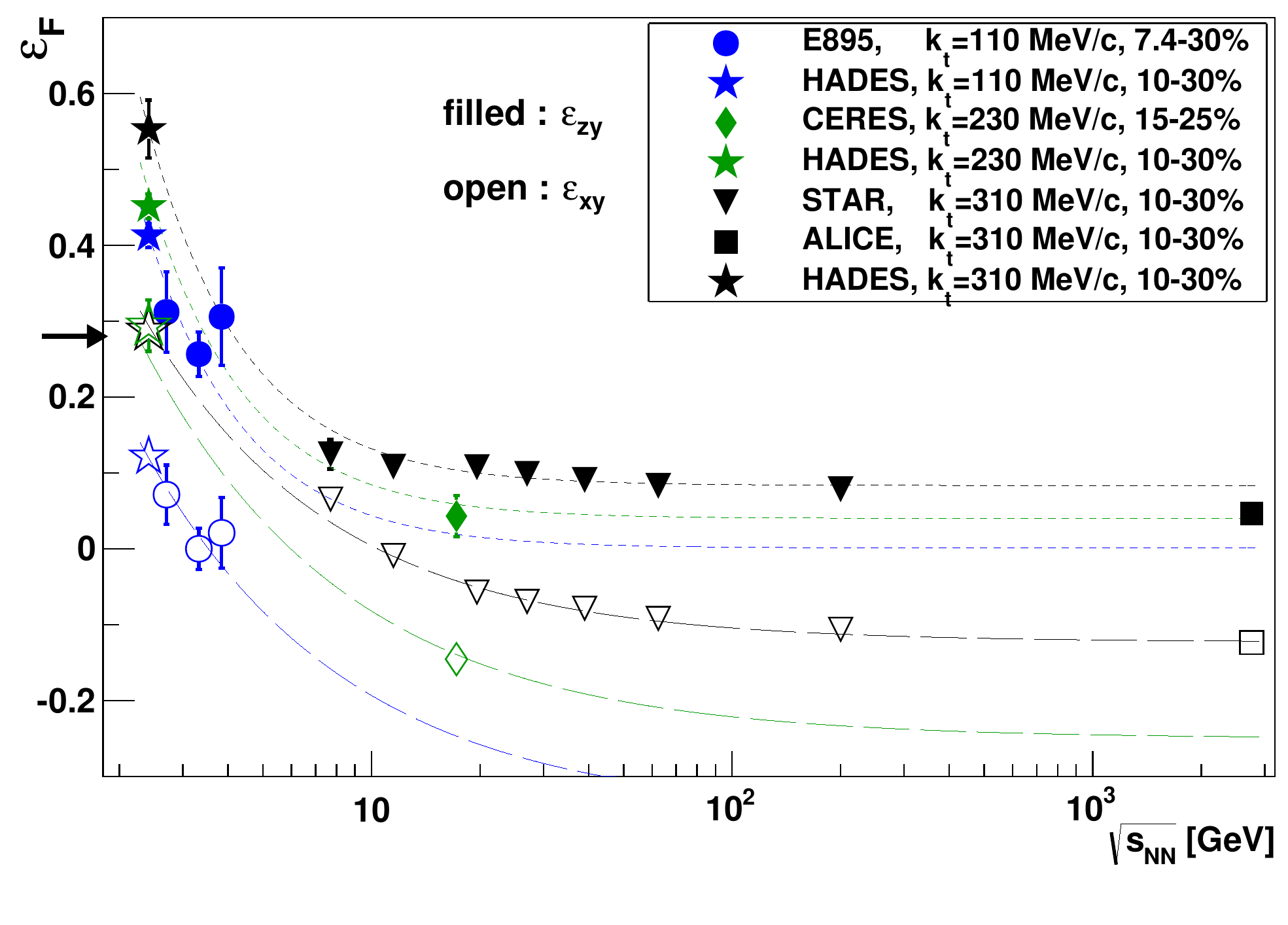}
\end{center}
\caption{Excitation function of the final $\pimin\pimin$ eccentricities  
$\epsilon_{xy}$ (open symbols) and $\epsilon_{zy}$ (filled symbols) for mid-central 
($10-30\,\%$) Au\,+\,Au, Pb\,+\,Au, or Pb\,+\,Pb collisions. Stars are HADES data 
for three different transverse momentum intervals with average values of 
$\bar k_\mathrm{t}= 110$, 230, and 310$~\mevc$. The circles, diamond, triangles, and square   
are corresponding data by E895 at AGS \cite{e895_2000_phi}, CERES at SPS \cite{CERES2008}, 
STAR at RHIC \cite{STAR2015}, and ALICE at LHC \cite{ALICE2017}, respectively. 
Error bars include only statistical uncertainties. Short (long) dashed curves connecting the 
filled (open) symbols are to guide the eyes. The arrow (to be compared to the open symbols) 
indicates the initial eccentricity derived from Glauber simulations 
\cite{PhobosGlauber2015}. 
}
\label{fig:pipi_ecc_excfct}
\end{figure}
\begin{figure}[h]
\begin{center}
\includegraphics[width=0.9\linewidth]{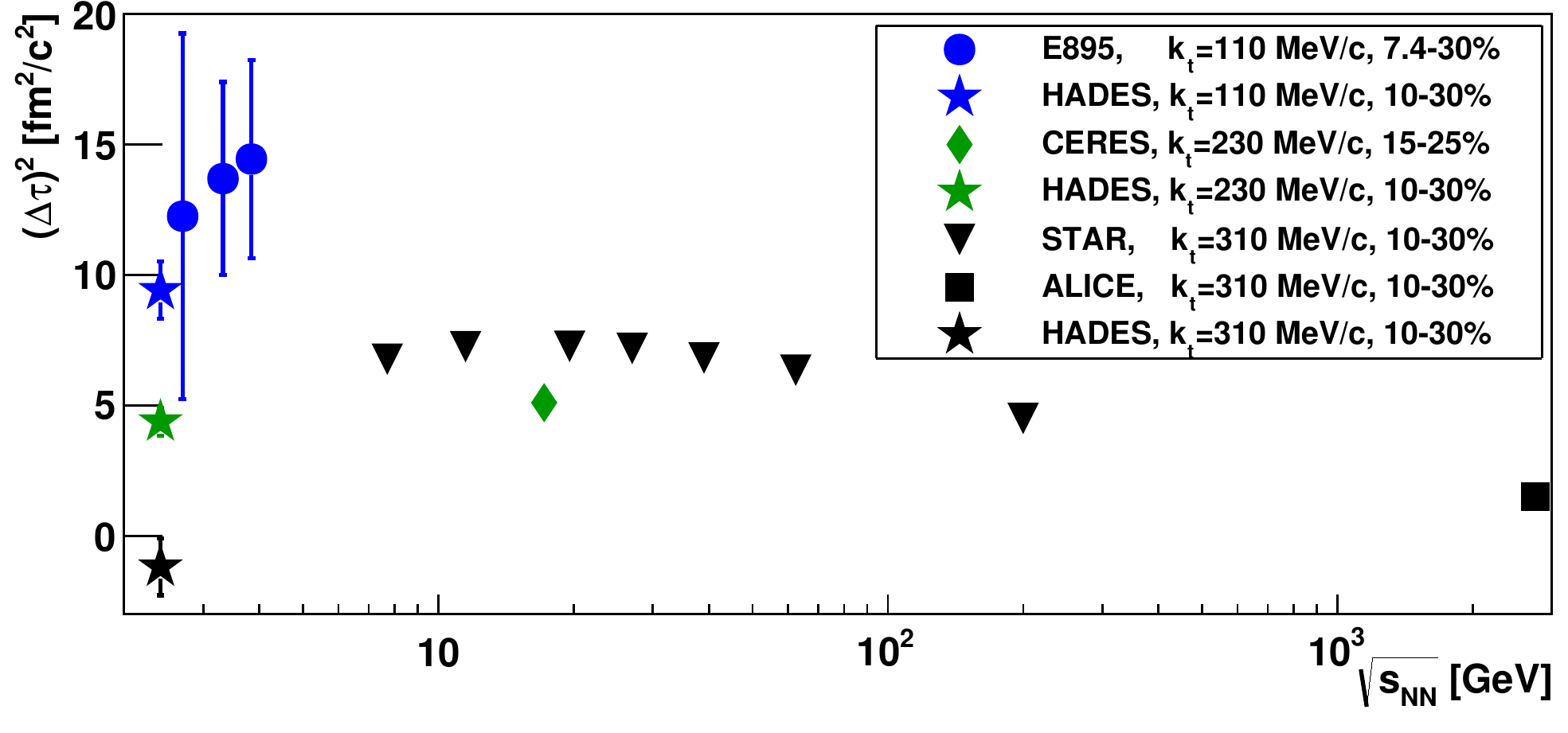}
\end{center}
\caption{Excitation function of the squared emission duration, the latter one 
either directly taken from  
the corresponding fit parameter of the spatial correlation tensor, $\sigma_\mathrm{t}^2=S_{00}$ 
(cf. Eqs.\,(\ref{R2_azi})), or derived from the difference of the 'out' and 'side' radii, 
$(\Delta\tau)^2 = (R^2_\mathrm{out}-R^2_\mathrm{side})/\langle\beta_t^2\rangle$, 
for mid-central ($10-30\,\%$) Au\,+\,Au collisions. Stars are HADES $\pimin\pimin$ data of 
$\sigma_\mathrm{t}^2$ for three different transverse momentum regions with average values of 
$\bar k_\mathrm{t}= 110$, 230, and 310$~\mevc$. The circles are corresponding 
$\sigma_\mathrm{t}^2$ data by E895 at AGS \cite{e895_2000_phi}. The diamond, triangles, and 
square represent data of $(\Delta\tau)^2$ by CERES at SPS \cite{CERES2008}, 
STAR at RHIC \cite{STAR2015}, and ALICE at LHC \cite{ALICE2017}, 
respectively. Error bars include only statistical uncertainties. }
\label{fig:dtau_exc_fct}
\end{figure}
\begin{figure}[h]
\begin{center}
\includegraphics[width=0.9\linewidth]{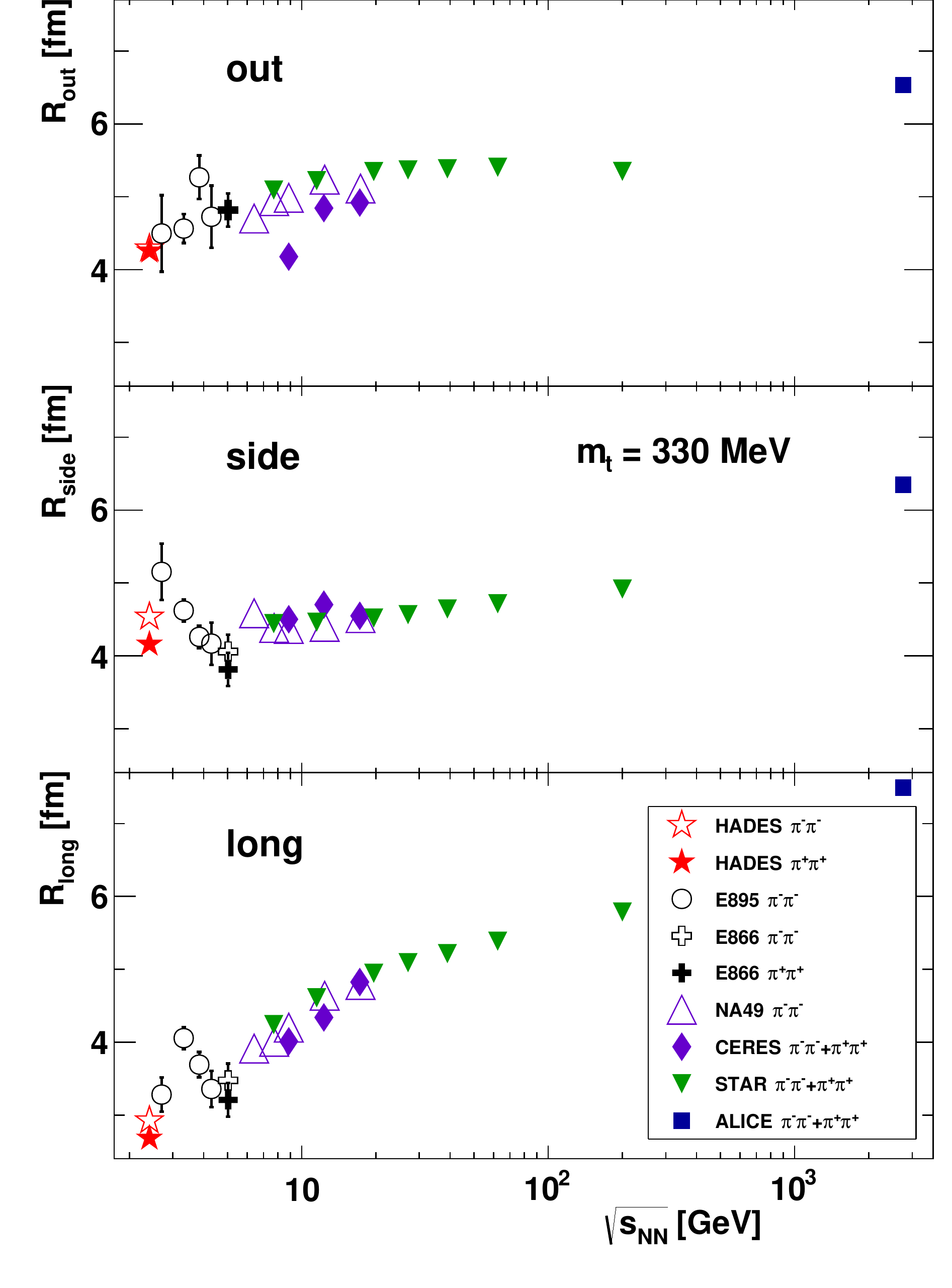}
\end{center}
\caption{Excitation function of the source radii $R_{\mathrm{out}}$ (upper panel), 
$R_{\mathrm{side}}$ (central panel), and 
$R_{\mathrm{long}}$ (lower panel) for the azimuthally-integrated correlation function of 
pairs of identical pions with transverse mass 
of $m_{t}=330~\mev$ in central ($0-10\,\%$) collisions of Au\,+\,Au or Pb\,+\,Pb.  
Squares represent data by ALICE at LHC ($\pi^+\pi^+$) \cite{ALICE2011}, 
full triangles STAR at RHIC ($\pi^-\pi^-$+$\pi^+\pi^+$) \cite{STAR2015}, diamonds are for CERES 
at SPS ($\pi^-\pi^-$+$\pi^+\pi^+$) \cite{CERES_2003}, open triangles are 
for NA49 at SPS ($\pi^-\pi^-$) \cite{NA49_2008}, open circles are $\pi^-\pi^-$ data by E895 
at AGS \cite{e895_2000,pipi_imaging_E895}, and open (full) crosses involve $\pi^-\pi^-$ 
($\pi^+\pi^+$) data of E866 at AGS \cite{e866_1999}, respectively. The present 
data of HADES at SIS18 for pairs of $\pi^-\pi^-$ ($\pi^+\pi^+$) are given as 
open (full) stars. Statistical errors are displayed as error bars; if not visible, they are 
smaller than the corresponding symbols. Note the suppressed zero on the ordinate.}
\label{fig:Rosl_exctfct_mt260_central_pimpim_pi0pi0}
\end{figure}
\begin{figure}[h]
\begin{center}
\includegraphics[width=0.9\linewidth]{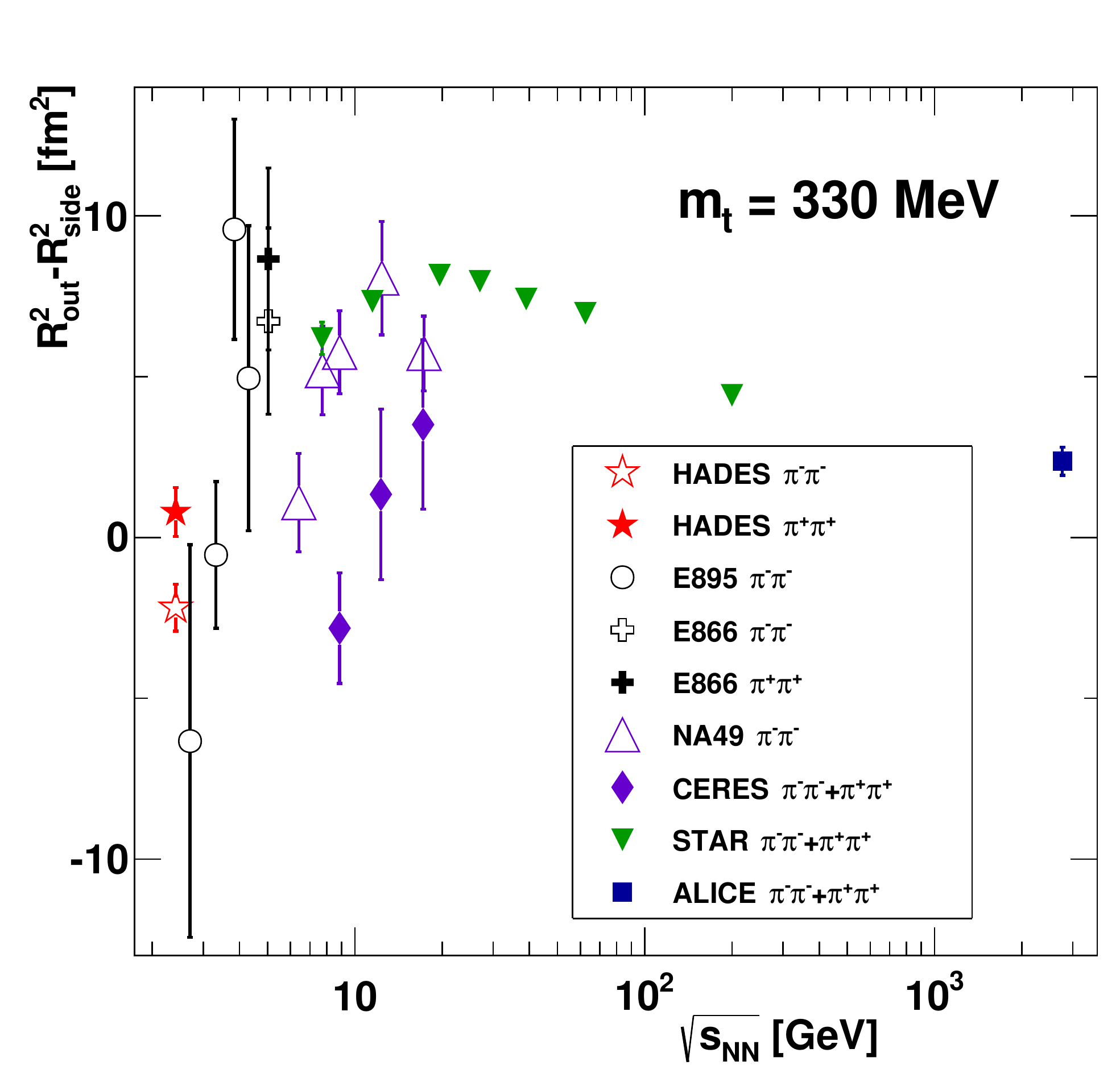}
\end{center}
\caption{The same as Fig.\,\ref{fig:Rosl_exctfct_mt260_central_pimpim_pi0pi0}, but for 
the derived quantity $R_{\mathrm{out}}^2-R_{\mathrm{side}}^2$. }
\label{fig:Rout2_Rside2_mt260_central_pimpim_pi0pi0}
\end{figure}
\begin{figure}[h]
\begin{center}
\includegraphics[width=0.9\linewidth]{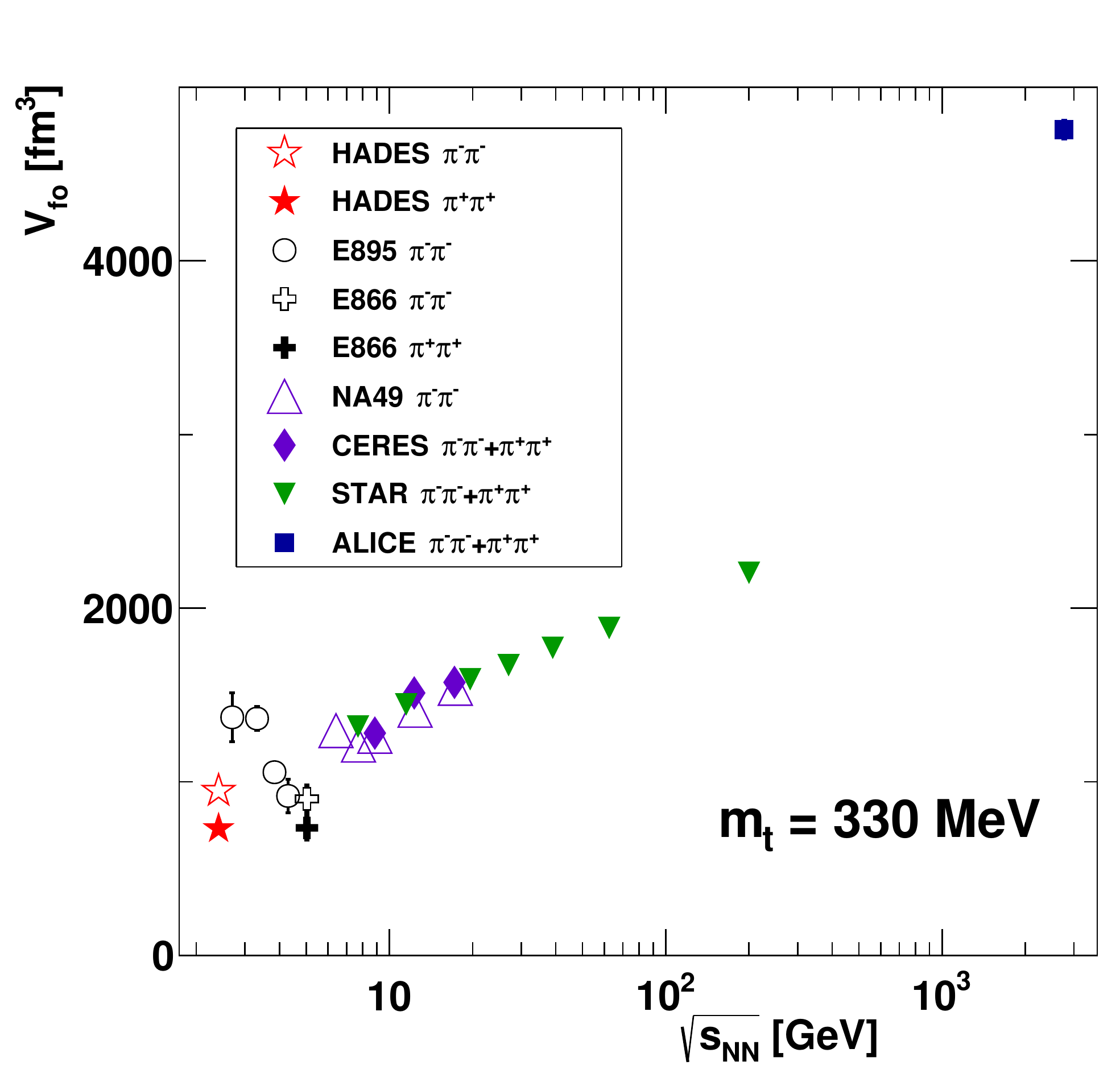}
\end{center}
\caption{The same as Fig.\,\ref{fig:Rosl_exctfct_mt260_central_pimpim_pi0pi0}, but for the 
derived approximate volume of the region of homogeneity, Eq.\,(\ref{Vfo_2RsideRlong}). }
\label{fig:Vfreezeout_mt160_central_pimpim_pi0pi0}
\end{figure}
%
%
\section{Summary and outlook}
\label{sect:summary}
We presented high-statistics $\pimin\pimin$ and $\piplus\piplus$ HBT data for \AuAu. 
The three-dimensional Gaussian emission source is studied in dependence on transverse 
momentum and collision centrality. It is found to follow the trends observed at higher 
collision energies, extending the corresponding excitation functions towards very low energies. 
A surprisingly small variation of the space-time extent of the 
pion emission source over three orders of magnitude of $\sqrt{s_{\mathrm{NN}}}$ is observed. 
All source radii increase almost linearly with the number of participants, 
irrespective of transverse momentum. Substantial differences of the source radii for pairs of 
negatively and positively charged pions, especially at low transverse momenta, are found, 
an effect hardly visible at higher collision energies. Correcting for this Coulomb effect,  
we found for all transverse momenta and centralities 
a stable hierarchy of the three widths of the emission ellipsoid,  
revealing an oblate shape in coordinate space with 
the largest extent oriented perpendicular to the 
reaction-plane. The corresponding eccentricity 
is found to recover the initial geometrical eccentricity of the nucleons if sufficiently 
large transverse momenta of the pions are considered. For low momenta, the shape 
approaches a circular one. Selecting collisions of the 30\,\% most central event classes, 
the tilt angle in the reaction plane, $\theta_\mathrm{s}$, is found to approach zero at 
high $p_\mathrm{t}$. For low transverse momenta, $\vert\theta_\mathrm{s}\vert$ increases and 
its $p_{\mathrm{t}}$ dependence is stronger for larger impact parameters. 
The centrality dependences of the tilt angle and 
of the eccentricity are found to be independent of pion charge and transverse momentum.  
Both quantities fit well into a corresponding 
low-energy excitation function for semi-central collisions \cite{e895_2000_phi}.

Future femtoscopic studies of other system-energy combinations in the 
'low-energy domain', e.g. performed 
with HADES\,+\,CBM  at FAIR/GSI \cite{CBM2017} and MPD at NICA in Dubna \cite{NICA} 
or with the STAR fixed-target program \cite{KMeehan_STAR_2017},  
may focus also on correlations of other particles (pp, p$\Lambda$, 
$\Lambda\Lambda$, etc.), allowing for deeper insight into their strong 
interaction \cite{CATS2018}. 

\begin{acknowledgement}
The HADES Collaboration gratefully acknowledges the support by the grants 
SIP JUC Cracow, Cracow (Poland), National Science Center, 2016/23/P/ST2/040 POLONEZ, 
2017/25/N/ST2/00580, 2017/26/M/ST2/00600; 
TU Darmstadt, Darmstadt (Germany), VH-NG-823, DFG GRK 2128, DFG CRC-TR 211, BMBF:05P18RDFC1; 
Goethe-University, Frankfurt (Germany) and TU Darmstadt, Darmstadt (Germany), ExtreMe Matter
Institute EMMI at GSI Darmstadt;
TU M\"unchen, Garching (Germany), MLL M\"unchen, DFG EClust 153, GSI TMLRG1316F, BMBF 05P15WOFCA, 
SFB 1258, DFG FAB898/2-2; NRNU MEPhI Moscow, Moscow (Russia), in framework of Russian 
Academic Excellence Project 02.a03.21.0005, Ministry of Science and Education of the Russian 
Federation 3.3380.2017/4.6; JLU Giessen, Giessen (Germany), BMBF:05P12RGGHM; IPN Orsay, 
Orsay Cedex (France), CNRS/IN2P3; NPI CAS, Rez, Rez (Czech Republic), 
MSMT LM2015049, OP VVV CZ.02.1.01/0.0/0.0/16 013/0001677, LTT17003. 
\end{acknowledgement}

%

\onecolumn
\newpage

\begin{table}[ht!]
\centering
\caption{Source parameters resulting from fits with Eqs.\,(\ref{pipi_fit_fct}) and 
(\ref{pipi_fit_fct_3dim}) for $\pi^-\pi^-$ pairs in dependence of centrality and average  
transverse momentum, $\bar k_\mathrm{t}$. Values in the 1st (2nd) brackets represent the 
corresponding statistical (systematic) 
uncertainties in units of the last digit of the respective quantity.}
\label{tab:pimpim}
\begin{tabular}{cc|cc|ccccc}
 \toprule
 Centrality & $\bar{k}_{\mathrm{t}}$  & $R_{\mathrm{inv}}$ & $\lambda_{\mathrm{inv}}$ & $R_{\mathrm{out}}$ & $R_{\mathrm{side}}$ & $R_{\mathrm{long}}$ & $\lambda_{\mathrm{osl}}$ \\
   (\%) & (MeV$/c$) & (fm) & & (fm) & (fm) & (fm) & \\
 \midrule
 0-10 & 43 & $7.03(23)(^{+17}_{-10})$ & $0.904(84)(^{+42}_{-38}) $ & $ 7.03(60)(^{+31}_{-8}) $ & $ 7.44(27)(^{+23}_{-0}) $ & $ 5.50(34)(^{+21}_{-2}) $ & $ 0.872(74)(^{+59}_{-0}) $ \\
 0-10 & 81 & $6.54(4)(^{+20}_{-0})$ & $0.801(12)(^{+30}_{-0}) $ & $ 6.88(9)(^{+17}_{-1}) $ & $ 7.15(10)(^{+16}_{-1}) $ & $ 5.53(5)(^{+11}_{-1}) $ & $ 0.904(19)(^{+18}_{-2}) $ \\
 0-10 & 125 & $5.88(2)(^{+28}_{-4})$ & $0.710(7)(^{+59}_{-6}) $ & $ 6.22(5)(^{+23}_{-1}) $ & $ 6.27(6)(^{+21}_{-1}) $ & $ 4.81(3)(^{+15}_{-0}) $ & $ 0.848(11)(^{+36}_{-1}) $ \\
 0-10 & 172 & $5.37(2)(^{+28}_{-14})$ & $0.662(7)(^{+65}_{-21}) $ & $ 5.59(6)(^{+26}_{-8}) $ & $ 5.75(5)(^{+15}_{-7}) $ & $ 4.17(2)(^{+15}_{-5}) $ & $ 0.840(12)(^{+40}_{-13}) $ \\
 0-10 & 221 & $5.01(3)(^{+26}_{-17})$ & $0.628(9)(^{+72}_{-26}) $ & $ 5.16(10)(^{+18}_{-9}) $ & $ 5.26(6)(^{+13}_{-8}) $ & $ 3.61(3)(^{+12}_{-5}) $ & $ 0.847(19)(^{+39}_{-17}) $ \\
 0-10 & 271 & $4.57(4)(^{+22}_{-17})$ & $0.565(12)(^{+74}_{-24}) $ & $ 4.69(34)(^{+16}_{-13}) $ & $ 4.80(8)(^{+11}_{-10}) $ & $ 3.11(3)(^{+11}_{-6}) $ & $ 0.836(62)(^{+43}_{-24}) $ \\
 0-10 & 321 & $4.25(6)(^{+15}_{-15})$ & $0.530(18)(^{+67}_{-20}) $ & $ 4.27(11)(^{+12}_{-8}) $ & $ 4.32(9)(^{+8}_{-6}) $ & $ 2.80(5)(^{+8}_{-11}) $ & $ 0.846(26)(^{+15}_{-15}) $ \\
 0-10 & 370 & $3.91(10)(^{+35}_{-14})$ & $0.462(27)(^{+110}_{-19}) $ & $ 3.84(27)(^{+20}_{-12}) $ & $ 3.95(13)(^{+14}_{-8}) $ & $ 2.50(7)(^{+15}_{-5}) $ & $ 0.776(57)(^{+57}_{-20}) $ \\
 0-10 & 420 & $3.84(15)(^{+28}_{-11})$ & $0.494(44)(^{+73}_{-15}) $ & $ 3.04(19)(^{+25}_{-4}) $ & $ 3.57(19)(^{+25}_{-12}) $ & $ 2.33(9)(^{+18}_{-0}) $ & $ 0.725(52)(^{+68}_{-3}) $ \\
 \midrule
 10-20 & 43 & $6.08(23)(^{+28}_{-15})$ & $0.734(75)(^{+82}_{-31}) $ & $ 4.82(73)(^{+69}_{-26}) $ & $ 6.24(24)(^{+14}_{-12}) $ & $ 5.73(39)(^{+37}_{-13}) $ & $ 0.740(72)(^{+30}_{-30}) $ \\
 10-20 & 81 & $6.16(4)(^{+9}_{-1})$ & $0.770(13)(^{+28}_{-1}) $ & $ 6.29(8)(^{+8}_{-0}) $ & $ 6.09(8)(^{+5}_{-0}) $ & $ 5.57(6)(^{+8}_{-0}) $ & $ 0.849(18)(^{+17}_{-0}) $ \\
 10-20 & 125 & $5.71(3)(^{+22}_{-2})$ & $0.702(9)(^{+50}_{-3}) $ & $ 5.94(5)(^{+17}_{-1}) $ & $ 5.73(5)(^{+11}_{-0}) $ & $ 4.89(3)(^{+11}_{-0}) $ & $ 0.829(12)(^{+26}_{-1}) $ \\
 10-20 & 172 & $5.21(3)(^{+22}_{-7})$ & $0.666(9)(^{+61}_{-10}) $ & $ 5.57(6)(^{+17}_{-2}) $ & $ 5.09(5)(^{+12}_{-2}) $ & $ 4.14(3)(^{+12}_{-1}) $ & $ 0.845(13)(^{+34}_{-3}) $ \\
 10-20 & 221 & $4.78(4)(^{+24}_{-13})$ & $0.629(12)(^{+73}_{-19}) $ & $ 5.06(10)(^{+17}_{-7}) $ & $ 4.66(6)(^{+7}_{-6}) $ & $ 3.60(3)(^{+8}_{-4}) $ & $ 0.868(20)(^{+31}_{-14}) $ \\
 10-20 & 270 & $4.32(5)(^{+25}_{-11})$ & $0.585(16)(^{+88}_{-17}) $ & $ 4.62(33)(^{+15}_{-11}) $ & $ 4.19(8)(^{+10}_{-7}) $ & $ 3.03(4)(^{+6}_{-4}) $ & $ 0.885(65)(^{+49}_{-19}) $ \\
 10-20 & 320 & $4.03(8)(^{+29}_{-10})$ & $0.529(25)(^{+100}_{-15}) $ & $ 4.11(13)(^{+23}_{-9}) $ & $ 3.62(10)(^{+30}_{-6}) $ & $ 2.73(6)(^{+11}_{-4}) $ & $ 0.827(31)(^{+81}_{-18}) $ \\
 10-20 & 370 & $3.52(14)(^{+60}_{-4})$ & $0.435(35)(^{+118}_{-5}) $ & $ 3.37(23)(^{+44}_{-15}) $ & $ 3.27(15)(^{+11}_{-4}) $ & $ 2.43(8)(^{+38}_{-2}) $ & $ 0.736(53)(^{+133}_{-10}) $ \\
 \midrule
 20-30 & 43 & $5.62(22)(^{+27}_{-13})$ & $0.821(86)(^{+56}_{-40}) $ & $ 4.22(74)(^{+118}_{-4}) $ & $ 5.49(22)(^{+27}_{-2}) $ & $ 5.76(44)(^{+55}_{-2}) $ & $ 0.790(80)(^{+108}_{-4}) $ \\
 20-30 & 81 & $5.69(4)(^{+16}_{-0})$ & $0.745(14)(^{+34}_{-0}) $ & $ 5.76(8)(^{+16}_{-0}) $ & $ 5.46(7)(^{+9}_{-0}) $ & $ 5.39(6)(^{+17}_{-0}) $ & $ 0.823(18)(^{+22}_{-0}) $ \\
 20-30 & 124 & $5.28(3)(^{+24}_{-2})$ & $0.701(10)(^{+48}_{-2}) $ & $ 5.56(5)(^{+20}_{-0}) $ & $ 5.06(5)(^{+13}_{-0}) $ & $ 4.62(4)(^{+16}_{-0}) $ & $ 0.822(12)(^{+29}_{-0}) $ \\
 20-30 & 172 & $4.87(3)(^{+22}_{-9})$ & $0.670(10)(^{+62}_{-13}) $ & $ 5.37(6)(^{+16}_{-3}) $ & $ 4.53(5)(^{+13}_{-2}) $ & $ 3.89(3)(^{+12}_{-2}) $ & $ 0.851(14)(^{+34}_{-4}) $ \\
 20-30 & 221 & $4.45(4)(^{+24}_{-13})$ & $0.622(14)(^{+81}_{-19}) $ & $ 4.84(10)(^{+10}_{-6}) $ & $ 4.21(6)(^{+9}_{-4}) $ & $ 3.32(4)(^{+5}_{-3}) $ & $ 0.861(21)(^{+16}_{-10}) $ \\
 20-30 & 270 & $4.03(7)(^{+34}_{-17})$ & $0.543(20)(^{+104}_{-24}) $ & $ 4.68(35)(^{+22}_{-9}) $ & $ 3.65(8)(^{+16}_{-5}) $ & $ 2.80(5)(^{+10}_{-3}) $ & $ 0.867(66)(^{+65}_{-16}) $ \\
 20-30 & 320 & $3.98(10)(^{+21}_{-9})$ & $0.565(34)(^{+72}_{-14}) $ & $ 4.13(16)(^{+25}_{-18}) $ & $ 3.42(12)(^{+15}_{-9}) $ & $ 2.58(7)(^{+4}_{-5}) $ & $ 0.868(40)(^{+60}_{-23}) $ \\
 20-30 & 370 & $3.57(18)(^{+73}_{-9})$ & $0.480(50)(^{+180}_{-12}) $ & $ 3.74(31)(^{+38}_{-18}) $ & $ 3.03(15)(^{+7}_{-0}) $ & $ 2.41(10)(^{+3}_{-8}) $ & $ 0.932(82)(^{+9}_{-6}) $ \\
 \midrule
 30-40 & 43 & $5.02(23)(^{+26}_{-9})$ & $0.711(89)(^{+42}_{-14}) $ & $ 3.74(45)(^{+17}_{-12}) $ & $ 5.18(23)(^{+12}_{-0}) $ & $ 4.96(58)(^{+44}_{-9}) $ & $ 0.721(86)(^{+42}_{-10}) $ \\
 30-40 & 81 & $5.10(5)(^{+26}_{-0})$ & $0.707(16)(^{+36}_{-0}) $ & $ 5.22(8)(^{+18}_{-2}) $ & $ 4.85(7)(^{+12}_{-2}) $ & $ 4.98(7)(^{+24}_{-2}) $ & $ 0.784(19)(^{+20}_{-3}) $ \\
 30-40 & 125 & $4.89(3)(^{+24}_{-1})$ & $0.691(12)(^{+41}_{-1}) $ & $ 5.26(6)(^{+19}_{-0}) $ & $ 4.53(5)(^{+9}_{-0}) $ & $ 4.34(4)(^{+15}_{-0}) $ & $ 0.811(14)(^{+20}_{-0}) $ \\
 30-40 & 172 & $4.52(4)(^{+27}_{-6})$ & $0.664(13)(^{+73}_{-10}) $ & $ 5.10(7)(^{+19}_{-3}) $ & $ 4.11(5)(^{+19}_{-2}) $ & $ 3.64(4)(^{+17}_{-2}) $ & $ 0.849(16)(^{+44}_{-4}) $ \\
 30-40 & 221 & $4.20(5)(^{+24}_{-12})$ & $0.648(17)(^{+81}_{-18}) $ & $ 4.64(10)(^{+13}_{-3}) $ & $ 3.72(6)(^{+16}_{-2}) $ & $ 3.07(4)(^{+6}_{-2}) $ & $ 0.872(23)(^{+17}_{-6}) $ \\
 30-40 & 270 & $3.81(8)(^{+37}_{-18})$ & $0.562(25)(^{+129}_{-26}) $ & $ 4.24(28)(^{+41}_{-7}) $ & $ 3.24(9)(^{+15}_{-4}) $ & $ 2.66(6)(^{+11}_{-2}) $ & $ 0.851(58)(^{+75}_{-11}) $ \\
 30-40 & 320 & $3.38(13)(^{+61}_{-8})$ & $0.480(36)(^{+155}_{-11}) $ & $ 4.03(19)(^{+30}_{-18}) $ & $ 2.86(12)(^{+38}_{-8}) $ & $ 2.37(8)(^{+11}_{-12}) $ & $ 0.855(47)(^{+39}_{-23}) $ \\
 \bottomrule
\end{tabular}

\end{table}

\begin{table}[ht!]
\centering
\caption{The same as table\,\ref{tab:pimpim}, but for $\pi^+\pi^+$ pairs.}
\label{tab:pippip}
\begin{tabular}{cc|cc|ccccc}
 \toprule
 Centrality & $\bar{k}_{\mathrm{t}}$  & $R_{\mathrm{inv}}$ & $\lambda_{\mathrm{inv}}$ & $R_{\mathrm{out}}$ & $R_{\mathrm{side}}$ & $R_{\mathrm{long}}$ & $\lambda_{\mathrm{osl}}$ \\
   (\%) & (MeV$/c$) & (fm) & & (fm) & (fm) & (fm) & \\
 \midrule
 0-10 & 92 & $4.81(16)(^{+39}_{-0})$ & $0.870(65)(^{+80}_{-0}) $ & $ 5.67(40)(^{+51}_{-3}) $ & $ 5.45(24)(^{+22}_{-4}) $ & $ 4.04(17)(^{+26}_{-2}) $ & $ 0.973(71)(^{+23}_{-5}) $ \\
 0-10 & 130 & $4.70(4)(^{+39}_{-5})$ & $0.697(14)(^{+81}_{-7}) $ & $ 5.32(7)(^{+40}_{-1}) $ & $ 5.05(7)(^{+20}_{-1}) $ & $ 3.87(4)(^{+22}_{-1}) $ & $ 0.876(17)(^{+40}_{-1}) $ \\
 0-10 & 175 & $4.54(3)(^{+31}_{-14})$ & $0.614(10)(^{+89}_{-20}) $ & $ 5.23(7)(^{+36}_{-8}) $ & $ 4.79(6)(^{+16}_{-7}) $ & $ 3.39(3)(^{+20}_{-4}) $ & $ 0.830(14)(^{+60}_{-12}) $ \\
 0-10 & 223 & $4.37(4)(^{+36}_{-17})$ & $0.556(11)(^{+104}_{-23}) $ & $ 4.86(9)(^{+26}_{-11}) $ & $ 4.62(6)(^{+14}_{-9}) $ & $ 3.13(3)(^{+13}_{-6}) $ & $ 0.815(19)(^{+43}_{-17}) $ \\
 0-10 & 272 & $4.17(5)(^{+34}_{-19})$ & $0.525(15)(^{+101}_{-25}) $ & $ 4.44(9)(^{+40}_{-9}) $ & $ 4.28(8)(^{+15}_{-7}) $ & $ 2.82(4)(^{+14}_{-4}) $ & $ 0.809(22)(^{+65}_{-15}) $ \\
 0-10 & 321 & $3.86(7)(^{+37}_{-17})$ & $0.489(19)(^{+111}_{-23}) $ & $ 4.16(12)(^{+30}_{-17}) $ & $ 4.06(9)(^{+10}_{-9}) $ & $ 2.55(5)(^{+8}_{-5}) $ & $ 0.835(28)(^{+66}_{-21}) $ \\
 0-10 & 371 & $3.57(10)(^{+35}_{-12})$ & $0.440(26)(^{+71}_{-14}) $ & $ 3.71(17)(^{+65}_{-27}) $ & $ 3.78(12)(^{+22}_{-7}) $ & $ 2.23(6)(^{+11}_{-6}) $ & $ 0.817(37)(^{+114}_{-25}) $ \\
 0-10 & 421 & $3.46(17)(^{+26}_{-9})$ & $0.376(37)(^{+54}_{-9}) $ & $ 3.28(22)(^{+48}_{-24}) $ & $ 3.55(19)(^{+10}_{-12}) $ & $ 2.12(9)(^{+4}_{-14}) $ & $ 0.723(48)(^{+17}_{-0}) $ \\
 \midrule
 10-20 & 92 & $4.86(18)(^{+9}_{-3})$ & $0.851(70)(^{+25}_{-6}) $ & $ 5.47(46)(^{+11}_{-6}) $ & $ 5.07(21)(^{+2}_{-15}) $ & $ 4.48(24)(^{+15}_{-5}) $ & $ 0.963(83)(^{+28}_{-10}) $ \\
 10-20 & 129 & $4.63(5)(^{+33}_{-5})$ & $0.678(16)(^{+77}_{-6}) $ & $ 5.32(8)(^{+24}_{-2}) $ & $ 4.52(7)(^{+11}_{-1}) $ & $ 3.88(5)(^{+22}_{-1}) $ & $ 0.836(19)(^{+42}_{-2}) $ \\
 10-20 & 174 & $4.53(4)(^{+26}_{-10})$ & $0.635(13)(^{+71}_{-14}) $ & $ 5.08(7)(^{+24}_{-5}) $ & $ 4.42(6)(^{+14}_{-4}) $ & $ 3.54(4)(^{+15}_{-3}) $ & $ 0.833(16)(^{+50}_{-6}) $ \\
 10-20 & 222 & $4.31(5)(^{+27}_{-15})$ & $0.587(15)(^{+84}_{-20}) $ & $ 4.78(10)(^{+21}_{-6}) $ & $ 4.04(6)(^{+16}_{-4}) $ & $ 3.14(4)(^{+10}_{-3}) $ & $ 0.828(21)(^{+43}_{-9}) $ \\
 10-20 & 271 & $4.05(6)(^{+22}_{-15})$ & $0.549(19)(^{+76}_{-20}) $ & $ 4.38(11)(^{+28}_{-8}) $ & $ 3.87(8)(^{+11}_{-6}) $ & $ 2.80(5)(^{+10}_{-4}) $ & $ 0.837(26)(^{+42}_{-15}) $ \\
 10-20 & 321 & $3.71(10)(^{+43}_{-8})$ & $0.477(27)(^{+111}_{-11}) $ & $ 4.05(14)(^{+4}_{-16}) $ & $ 3.49(11)(^{+9}_{-4}) $ & $ 2.40(6)(^{+3}_{-2}) $ & $ 0.796(33)(^{+25}_{-10}) $ \\
 10-20 & 371 & $3.80(16)(^{+25}_{-28})$ & $0.474(44)(^{+41}_{-35}) $ & $ 4.51(33)(^{+34}_{-73}) $ & $ 3.38(16)(^{+0}_{-14}) $ & $ 2.35(9)(^{+16}_{-6}) $ & $ 0.951(63)(^{+121}_{-34}) $ \\
 \midrule
 20-30 & 92 & $4.29(19)(^{+45}_{-1})$ & $0.721(67)(^{+91}_{-1}) $ & $ 4.71(54)(^{+23}_{-1}) $ & $ 4.34(21)(^{+15}_{-1}) $ & $ 4.01(25)(^{+19}_{-1}) $ & $ 0.774(76)(^{+34}_{-0}) $ \\
 20-30 & 129 & $4.36(6)(^{+39}_{-2})$ & $0.653(18)(^{+88}_{-2}) $ & $ 5.10(9)(^{+24}_{-0}) $ & $ 4.11(7)(^{+7}_{-0}) $ & $ 3.85(6)(^{+18}_{-0}) $ & $ 0.825(21)(^{+30}_{-0}) $ \\
 20-30 & 174 & $4.23(5)(^{+30}_{-9})$ & $0.623(15)(^{+79}_{-12}) $ & $ 4.92(8)(^{+24}_{-5}) $ & $ 3.86(6)(^{+17}_{-3}) $ & $ 3.37(4)(^{+13}_{-3}) $ & $ 0.828(18)(^{+50}_{-7}) $ \\
 20-30 & 222 & $4.02(6)(^{+33}_{-15})$ & $0.571(18)(^{+94}_{-21}) $ & $ 4.70(11)(^{+21}_{-9}) $ & $ 3.68(7)(^{+20}_{-6}) $ & $ 3.01(4)(^{+8}_{-4}) $ & $ 0.843(25)(^{+41}_{-14}) $ \\
 20-30 & 271 & $3.98(9)(^{+37}_{-23})$ & $0.570(27)(^{+130}_{-34}) $ & $ 4.40(13)(^{+26}_{-4}) $ & $ 3.36(9)(^{+26}_{-2}) $ & $ 2.72(6)(^{+13}_{-2}) $ & $ 0.879(33)(^{+71}_{-6}) $ \\
 20-30 & 321 & $3.50(12)(^{+34}_{-7})$ & $0.497(34)(^{+52}_{-9}) $ & $ 4.26(19)(^{+12}_{-24}) $ & $ 3.14(12)(^{+6}_{-10}) $ & $ 2.48(8)(^{+5}_{-12}) $ & $ 0.920(48)(^{+20}_{-30}) $ \\
 20-30 & 370 & $3.15(19)(^{+50}_{-3})$ & $0.416(47)(^{+122}_{-2}) $ & $ 4.27(38)(^{+22}_{-86}) $ & $ 2.69(16)(^{+9}_{-18}) $ & $ 2.24(11)(^{+9}_{-18}) $ & $ 0.985(77)(^{+47}_{-75}) $ \\
 \midrule
 30-40 & 92 & $4.16(23)(^{+46}_{-16})$ & $0.767(86)(^{+132}_{-27}) $ & $ 5.33(67)(^{+39}_{-7}) $ & $ 3.84(21)(^{+12}_{-6}) $ & $ 4.46(29)(^{+21}_{-6}) $ & $ 0.936(106)(^{+32}_{-11}) $ \\
 30-40 & 128 & $4.09(6)(^{+33}_{-5})$ & $0.659(21)(^{+73}_{-5}) $ & $ 4.68(10)(^{+25}_{-0}) $ & $ 3.75(7)(^{+4}_{-0}) $ & $ 3.73(7)(^{+24}_{-0}) $ & $ 0.816(25)(^{+32}_{-0}) $ \\
 30-40 & 173 & $3.95(6)(^{+44}_{-10})$ & $0.621(18)(^{+115}_{-13}) $ & $ 4.52(8)(^{+17}_{-1}) $ & $ 3.50(6)(^{+15}_{-1}) $ & $ 3.22(5)(^{+18}_{-0}) $ & $ 0.820(21)(^{+30}_{-0}) $ \\
 30-40 & 222 & $3.86(7)(^{+31}_{-14})$ & $0.592(23)(^{+89}_{-20}) $ & $ 4.59(12)(^{+5}_{-5}) $ & $ 3.26(7)(^{+9}_{-3}) $ & $ 2.79(5)(^{+6}_{-2}) $ & $ 0.858(29)(^{+13}_{-7}) $ \\
 30-40 & 271 & $3.54(11)(^{+45}_{-22})$ & $0.505(31)(^{+129}_{-29}) $ & $ 4.17(15)(^{+32}_{-8}) $ & $ 3.07(11)(^{+37}_{-5}) $ & $ 2.46(7)(^{+9}_{-2}) $ & $ 0.808(38)(^{+73}_{-9}) $ \\
 30-40 & 321 & $3.44(15)(^{+20}_{-9})$ & $0.517(47)(^{+58}_{-11}) $ & $ 3.76(22)(^{+5}_{-13}) $ & $ 2.94(15)(^{+0}_{-42}) $ & $ 2.34(10)(^{+10}_{-16}) $ & $ 0.867(59)(^{+19}_{-40}) $ \\
 \bottomrule
\end{tabular}

\end{table}

\begin{table}[ht!]
\centering
\caption{The same as table\,\ref{tab:pimpim}, but for constructed $\tilde{\pi}^0\tilde{\pi}^0$ pairs. In 
addition, the Coulomb potentials extracted from the $\pi^-\pi^-$ and $\pi^+\pi^+$ data 
according to the method described in Sect.\,\ref{sect:pi0pi0_corr} are 
given in columns 4 and $8-10$.}
\label{tab:pi0pi0}
\begin{tabular}{cc|cc|ccccccc}
 \toprule
 Centrality & $\bar{k}_{\mathrm{t}}$  & $R_{\mathrm{inv}}$ & $V^{\mathrm{eff}}_{\mathrm{inv}}$ & $R_{\mathrm{out}}$ & $R_{\mathrm{side}}$ & $R_{\mathrm{long}}$ & $V^{\mathrm{eff}}_{\mathrm{out}}$ & $V^{\mathrm{eff}}_{\mathrm{side}}$ & $V^{\mathrm{eff}}_{\mathrm{long}}$ \\
   (\%) & (MeV$/c$) & (fm) & (MeV) & (fm) & (fm) & (fm) & (MeV) & (MeV) & (MeV) \\
 \midrule
 0-10 & 87 & 5.70(4) & 7.0(5) & 6.30(27) & 6.32(18) & 4.81(12) & 3.1(14) & 5.5(6) & 5.3(11) \\
 0-10 & 127 & 5.30(3) & 9.7(3) & 5.78(6) & 5.67(9) & 4.35(5) & 5.5(6) & 9.1(5) & 7.3(4) \\
 0-10 & 174 & 4.97(3) & 10.9(4) & 5.41(6) & 5.29(8) & 3.80(4) & 3.8(9) & 11.7(6) & 13.0(4) \\
 0-10 & 222 & 4.70(4) & 11.0(7) & 5.01(10) & 4.94(9) & 3.38(4) & 2.8(21) & 11.0(10) & 11.6(7) \\
 0-10 & 271 & 4.38(6) & 9.0(13) & 4.57(26) & 4.55(11) & 2.97(5) & -0.1(61) & 12.8(16) & 10.4(13) \\
 0-10 & 321 & 4.06(8) & 13.1(24) & 4.21(11) & 4.19(13) & 2.68(7) & 2.1(69) & 9.2(27) & 12.4(24) \\
 0-10 & 371 & 3.75(13) & 16.9(47) & 3.77(23) & 3.87(18) & 2.37(9) & 5.0(92) & 10.3(49) & 17.6(48) \\
 0-10 & 421 & 3.55(21) & 26.5(87) & 3.24(20) & 3.55(26) & 2.15(14) & -3.2(155) & 1.4(91) & 27.6(93) \\
 \midrule
 10-20 & 87 & 5.54(4) & 5.0(6) & 5.92(31) & 5.63(16) & 5.07(16) & 0.7(18) & 3.7(7) & 2.3(13) \\
 10-20 & 127 & 5.19(3) & 8.5(4) & 5.63(7) & 5.15(8) & 4.40(6) & 4.4(7) & 9.5(5) & 7.7(4) \\
 10-20 & 173 & 4.88(4) & 9.2(5) & 5.33(7) & 4.76(8) & 3.85(5) & 5.0(10) & 9.0(6) & 9.3(5) \\
 10-20 & 221 & 4.55(5) & 9.5(9) & 4.92(10) & 4.36(9) & 3.38(5) & 5.4(22) & 12.8(10) & 12.2(9) \\
 10-20 & 271 & 4.19(8) & 7.4(17) & 4.50(25) & 4.03(11) & 2.92(6) & -0.6(63) & 8.5(19) & 8.0(17) \\
 10-20 & 321 & 3.88(13) & 12.0(35) & 4.08(14) & 3.56(14) & 2.57(9) & 5.3(77) & 7.5(34) & 21.5(35) \\
 10-20 & 370 & 3.66(21) & -13.1(71) & 3.98(30) & 3.33(22) & 2.39(13) & -11.5(121) & -4.0(66) & 17.9(69) \\
 \midrule
 20-30 & 86 & 5.02(5) & 6.2(7) & 5.25(35) & 4.93(15) & 4.71(17) & 2.6(23) & 5.0(7) & 4.7(17) \\
 20-30 & 127 & 4.83(4) & 8.0(4) & 5.32(7) & 4.60(8) & 4.24(7) & 3.5(8) & 8.4(5) & 6.1(5) \\
 20-30 & 173 & 4.56(5) & 9.3(6) & 5.15(7) & 4.21(7) & 3.64(5) & 6.1(11) & 10.6(7) & 8.7(6) \\
 20-30 & 221 & 4.24(7) & 9.5(11) & 4.77(10) & 3.95(9) & 3.17(6) & 0.8(24) & 12.3(12) & 7.4(11) \\
 20-30 & 271 & 4.01(9) & 1.4(23) & 4.54(27) & 3.50(12) & 2.76(7) & 0.8(67) & 9.0(22) & 1.0(23) \\
 20-30 & 320 & 3.75(15) & 18.9(43) & 4.19(18) & 3.28(16) & 2.53(11) & -3.4(93) & 13.3(42) & 2.2(46) \\
 20-30 & 370 & 3.37(25) & 21.7(96) & 4.01(35) & 2.86(22) & 2.33(14) & -8.8(142) & 22.1(76) & 22.5(79) \\
 \midrule
 30-40 & 86 & 4.65(6) & 4.5(10) & 5.37(47) & 4.35(15) & 4.76(21) & 2.3(33) & 4.7(8) & 0.3(27) \\
 30-40 & 126 & 4.51(5) & 7.3(5) & 4.98(8) & 4.15(8) & 4.04(8) & 5.3(10) & 7.4(5) & 5.1(7) \\
 30-40 & 173 & 4.24(6) & 8.9(8) & 4.81(8) & 3.81(8) & 3.43(6) & 6.5(14) & 10.1(8) & 7.0(8) \\
 30-40 & 221 & 4.03(8) & 7.7(15) & 4.61(11) & 3.49(9) & 2.94(7) & 1.8(28) & 12.0(13) & 7.2(15) \\
 30-40 & 271 & 3.68(12) & 8.8(31) & 4.21(22) & 3.15(14) & 2.56(9) & -6.7(71) & 6.3(27) & 4.7(29) \\
 30-40 & 321 & 3.41(23) & -2.7(61) & 3.90(21) & 2.90(19) & 2.36(13) & 10.4(109) & -6.2(52) & 11.9(61) \\
 \bottomrule
\end{tabular}

\end{table}

\begin{table}[ht!]
\centering
\caption{Parameters from the fits with Eq.\,(\ref{fit_R0_alpha}) to the data of 
Fig.\,\ref{fig:Radien_inv_osl_pi0pi0_allcentralities_newbinning_pt100to800}.} 
\label{tab:alpha_R0}
\begin{tabular}{c|cc|cccccc}
 \toprule
 Centrality & $\alpha_{\mathrm{inv}}$ & $R_{0,\,\mathrm{inv}}$ & $\alpha_{\mathrm{out}}$ & $R_{0,\,\mathrm{out}}$ & $\alpha_{\mathrm{side}}$ & $R_{0,\,\mathrm{side}}$ & $\alpha_{\mathrm{long}}$ & $R_{0,\,\mathrm{long}}$ \\
   (\%) & & (fm) & & (fm) & & (fm) & & (fm) \\
 \midrule
 0-10 & $-0.42(2)$ & $6.06(4)$ & $-0.49(4)$ & $6.75(13)$ & $-0.49(4)$ & $6.67(14)$ & $-0.78(3)$ & $5.50(9)$ \\
 10-20 & $-0.44(2)$ & $5.94(5)$ & $-0.47(4)$ & $6.55(14)$ & $-0.55(4)$ & $6.13(13)$ & $-0.86(4)$ & $5.74(11)$ \\
 20-30 & $-0.37(3)$ & $5.37(6)$ & $-0.34(5)$ & $5.93(15)$ & $-0.53(5)$ & $5.39(13)$ & $-0.87(4)$ & $5.47(13)$ \\
 30-40 & $-0.35(4)$ & $4.97(7)$ & $-0.33(6)$ & $5.56(16)$ & $-0.53(6)$ & $4.84(14)$ & $-0.96(6)$ & $5.38(16)$ \\
 \bottomrule
\end{tabular}

\end{table}

\begin{table}[ht!]
\centering
\caption{The geometrical and temporal variances (columns 3 to 6) and the 
tilt angle (column 7) of the $\pi^-\pi^-$ emission source in dependence of centrality 
and average transverse momentum, $\bar k_\mathrm{t}$. Values in the 1st (2nd) brackets 
represent the corresponding statistical (systematic) uncertainties 
in units of the last digit of the respective quantity.}
\label{tab:pimpim_azimuthal}
\begin{tabular}{cc|cccc|ccc|c}
 \toprule
 Centrality & $\bar{k}_{\mathrm{t}}$ & $\sigma^2_{\mathrm{x}}$ & $\sigma^2_{\mathrm{y}}$ & $\sigma^2_{\mathrm{z}}$ & $\sigma^2_{\mathrm{t}}$ & $\theta_{\mathrm{s}}$  \\
   (\%) & (MeV$/c$) & (fm$^2$) & (fm$^2$) & (fm$^2$) & (fm$^2$/c$^2$) & (deg) \\
 \midrule
 0-10 & 82 & $48.1(14)(^{+6}_{-7})$ & $55.4(16)(^{+15}_{-6}) $ & $ 23.9(8)(^{+3}_{-3}) $ & $ -11.2(35)(^{+14}_{-22}) $ & $ -27(2)(^{+0}_{-0}) $  \\
 0-10 & 126 & $38.1(9)(^{+18}_{-9})$ & $43.8(9)(^{+22}_{-7}) $ & $ 21.0(4)(^{+9}_{-4}) $ & $ -1.8(12)(^{+4}_{-6}) $ & $ -20(1)(^{+0}_{-1}) $  \\
 0-10 & 173 & $31.5(8)(^{+21}_{-12})$ & $38.8(8)(^{+25}_{-14}) $ & $ 17.0(2)(^{+10}_{-6}) $ & $ -3.2(8)(^{+1}_{-3}) $ & $ -12(1)(^{+0}_{-0}) $  \\
 0-10 & 222 & $26.2(9)(^{+17}_{-12})$ & $33.2(9)(^{+23}_{-15}) $ & $ 12.5(2)(^{+7}_{-5}) $ & $ -3.2(7)(^{+1}_{-1}) $ & $ -8(1)(^{+0}_{-0}) $  \\
 0-10 & 271 & $22.0(10)(^{+14}_{-13})$ & $29.5(11)(^{+20}_{-16}) $ & $ 9.0(1)(^{+5}_{-5}) $ & $ -9.2(7)(^{+3}_{-3}) $ & $ -5(1)(^{+0}_{-0}) $  \\
 \midrule
 10-20 & 82 & $41.8(7)(^{+10}_{-5})$ & $43.7(9)(^{+10}_{-7}) $ & $ 18.9(7)(^{+4}_{-4}) $ & $ 11.1(30)(^{+13}_{-19}) $ & $ -46(1)(^{+0}_{-0}) $  \\
 10-20 & 125 & $29.7(5)(^{+15}_{-7})$ & $39.5(7)(^{+19}_{-7}) $ & $ 18.0(4)(^{+9}_{-3}) $ & $ 7.0(12)(^{+3}_{-4}) $ & $ -40(2)(^{+0}_{-0}) $  \\
 10-20 & 173 & $23.0(5)(^{+13}_{-7})$ & $33.4(6)(^{+16}_{-9}) $ & $ 15.4(3)(^{+8}_{-4}) $ & $ 7.6(9)(^{+6}_{-4}) $ & $ -27(2)(^{+0}_{-0}) $  \\
 10-20 & 221 & $18.7(6)(^{+10}_{-9})$ & $30.0(7)(^{+15}_{-12}) $ & $ 11.4(2)(^{+6}_{-5}) $ & $ 2.5(8)(^{+4}_{-3}) $ & $ -14(2)(^{+0}_{-0}) $  \\
 \midrule
 20-30 & 82 & $33.5(6)(^{+7}_{-5})$ & $37.3(8)(^{+14}_{-6}) $ & $ 14.7(7)(^{+2}_{-5}) $ & $ 14.6(28)(^{+20}_{-27}) $ & $ -58(1)(^{+0}_{-0}) $  \\
 20-30 & 125 & $22.1(3)(^{+11}_{-5})$ & $35.0(6)(^{+15}_{-8}) $ & $ 13.6(5)(^{+6}_{-3}) $ & $ 11.9(11)(^{+8}_{-8}) $ & $ -60(2)(^{+0}_{-0}) $  \\
 20-30 & 172 & $15.2(4)(^{+10}_{-6})$ & $30.0(6)(^{+13}_{-9}) $ & $ 12.9(3)(^{+7}_{-5}) $ & $ 10.3(8)(^{+11}_{-8}) $ & $ -42(6)(^{+1}_{-4}) $  \\
 20-30 & 221 & $13.8(6)(^{+7}_{-7})$ & $26.7(7)(^{+13}_{-12}) $ & $ 9.8(2)(^{+5}_{-4}) $ & $ 3.5(8)(^{+7}_{-3}) $ & $ -7(3)(^{+0}_{-1}) $  \\
 \midrule
 25-35 & 82 & $29.1(6)(^{+8}_{-7})$ & $34.4(8)(^{+10}_{-5}) $ & $ 12.8(7)(^{+4}_{-3}) $ & $ 17.4(28)(^{+28}_{-13}) $ & $ -63(1)(^{+0}_{-0}) $  \\
 25-35 & 125 & $19.4(3)(^{+11}_{-5})$ & $31.0(6)(^{+17}_{-9}) $ & $ 12.1(5)(^{+6}_{-3}) $ & $ 15.2(11)(^{+13}_{-6}) $ & $ -66(2)(^{+0}_{-0}) $  \\
 25-35 & 172 & $12.7(2)(^{+9}_{-5})$ & $27.5(5)(^{+15}_{-9}) $ & $ 11.1(5)(^{+8}_{-5}) $ & $ 11.1(8)(^{+13}_{-8}) $ & $ -75(7)(^{+2}_{-2}) $  \\
 25-35 & 221 & $10.5(2)(^{+3}_{-5})$ & $22.8(7)(^{+8}_{-7}) $ & $ 12.0(6)(^{+4}_{-3}) $ & $ 6.6(9)(^{+9}_{-8}) $ & $ -94(7)(^{+3}_{-0}) $  \\
 \midrule
 30-45 & 82 & $27.0(6)(^{+1}_{-11})$ & $32.6(8)(^{+6}_{-8}) $ & $ 11.2(8)(^{+0}_{-5}) $ & $ 20.2(29)(^{+30}_{-43}) $ & $ -69(2)(^{+0}_{-1}) $  \\
 30-45 & 125 & $16.3(3)(^{+10}_{-3})$ & $26.9(5)(^{+15}_{-5}) $ & $ 10.6(5)(^{+6}_{-4}) $ & $ 15.5(11)(^{+13}_{-11}) $ & $ -75(2)(^{+0}_{-1}) $  \\
 30-45 & 172 & $11.1(2)(^{+7}_{-4})$ & $25.5(5)(^{+15}_{-8}) $ & $ 8.8(5)(^{+6}_{-4}) $ & $ 10.4(8)(^{+12}_{-6}) $ & $ -94(4)(^{+1}_{-1}) $  \\
 \bottomrule
\end{tabular}

\end{table}

\begin{table}[ht!]
\centering
\caption{The same as table\,\ref{tab:pimpim_azimuthal}, but for $\pi^+\pi^+$ pairs.}
\label{tab:pippip_azimuthal}
\begin{tabular}{cc|cccc|ccc|c}
 \toprule
 Centrality & $\bar{k}_{\mathrm{t}}$ & $\sigma^2_{\mathrm{x}}$ & $\sigma^2_{\mathrm{y}}$ & $\sigma^2_{\mathrm{z}}$ & $\sigma^2_{\mathrm{t}}$ & $\theta_{\mathrm{s}}$  \\
   (\%) & (MeV$/c$) & (fm$^2$) & (fm$^2$) & (fm$^2$) & (fm$^2$/c$^2$) & (deg) \\
 \midrule
 0-10 & 132 & $24.3(11)(^{+1}_{-5})$ & $30.5(12)(^{+35}_{-7}) $ & $ 12.9(4)(^{+2}_{-4}) $ & $ 2.9(15)(^{+5}_{-22}) $ & $ -18(2)(^{+0}_{-4}) $  \\
 0-10 & 175 & $22.2(8)(^{+20}_{-14})$ & $26.8(9)(^{+23}_{-13}) $ & $ 10.9(2)(^{+9}_{-6}) $ & $ 4.5(9)(^{+7}_{-8}) $ & $ -12(1)(^{+0}_{-1}) $  \\
 0-10 & 223 & $22.7(8)(^{+18}_{-14})$ & $21.9(8)(^{+18}_{-13}) $ & $ 8.8(1)(^{+8}_{-6}) $ & $ 0.3(7)(^{+5}_{-4}) $ & $ -6(1)(^{+0}_{-0}) $  \\
 0-10 & 272 & $17.1(9)(^{+10}_{-9})$ & $21.6(9)(^{+17}_{-10}) $ & $ 7.4(2)(^{+6}_{-4}) $ & $ -2.8(7)(^{+5}_{-3}) $ & $ -6(1)(^{+0}_{-1}) $  \\
 \midrule
 10-20 & 131 & $22.2(7)(^{+14}_{-9})$ & $27.3(9)(^{+16}_{-10}) $ & $ 11.9(5)(^{+6}_{-6}) $ & $ 9.5(16)(^{+9}_{-10}) $ & $ -34(2)(^{+1}_{-0}) $  \\
 10-20 & 175 & $17.8(6)(^{+12}_{-5})$ & $25.0(7)(^{+14}_{-6}) $ & $ 10.2(3)(^{+7}_{-3}) $ & $ 8.1(10)(^{+7}_{-3}) $ & $ -24(2)(^{+0}_{-1}) $  \\
 10-20 & 223 & $14.1(6)(^{+7}_{-4})$ & $25.0(7)(^{+11}_{-6}) $ & $ 8.9(2)(^{+4}_{-2}) $ & $ 3.1(8)(^{+7}_{-2}) $ & $ -15(2)(^{+0}_{-0}) $  \\
 \midrule
 20-30 & 131 & $18.2(7)(^{+10}_{-2})$ & $25.1(9)(^{+17}_{-7}) $ & $ 10.0(7)(^{+5}_{-8}) $ & $ 12.5(20)(^{+18}_{-24}) $ & $ -50(3)(^{+1}_{-2}) $  \\
 20-30 & 175 & $14.0(5)(^{+7}_{-4})$ & $22.6(6)(^{+9}_{-5}) $ & $ 9.7(4)(^{+4}_{-3}) $ & $ 9.4(10)(^{+11}_{-8}) $ & $ -32(4)(^{+1}_{-1}) $  \\
 20-30 & 238 & $11.3(5)(^{+3}_{-4})$ & $19.7(6)(^{+7}_{-7}) $ & $ 8.5(2)(^{+2}_{-3}) $ & $ 7.5(8)(^{+8}_{-5}) $ & $ -13(4)(^{+1}_{-1}) $  \\
 \midrule
 25-35 & 131 & $18.6(7)(^{+13}_{-5})$ & $22.8(9)(^{+11}_{-4}) $ & $ 9.7(9)(^{+4}_{-5}) $ & $ 13.7(23)(^{+33}_{-21}) $ & $ -61(3)(^{+3}_{-1}) $  \\
 25-35 & 175 & $13.7(4)(^{+8}_{-10})$ & $22.8(7)(^{+11}_{-11}) $ & $ 11.1(6)(^{+5}_{-7}) $ & $ 9.5(11)(^{+14}_{-16}) $ & $ -56(7)(^{+3}_{-3}) $  \\
 25-35 & 193 & $10.2(3)(^{+4}_{-3})$ & $21.1(5)(^{+9}_{-4}) $ & $ 8.6(3)(^{+2}_{-3}) $ & $ 8.9(7)(^{+11}_{-7}) $ & $ -41(8)(^{+1}_{-4}) $  \\
 \midrule
 30-45 & 158 & $12.7(3)(^{+3}_{-7})$ & $21.6(5)(^{+4}_{-8}) $ & $ 7.3(4)(^{+1}_{-4}) $ & $ 12.4(9)(^{+13}_{-14}) $ & $ -69(2)(^{+0}_{-0}) $  \\
 30-45 & 193 & $10.7(2)(^{+5}_{-8})$ & $22.1(6)(^{+9}_{-11}) $ & $ 8.0(5)(^{+3}_{-4}) $ & $ 11.5(9)(^{+11}_{-12}) $ & $ -79(4)(^{+0}_{-0}) $  \\
 \bottomrule
\end{tabular}

\end{table}

\end{document}